%% file: manuscript.tex
\documentclass[letterpaper, 10 pt, conference]{ieeeconf}  
\IEEEoverridecommandlockouts                              
\newif\ifextendedversion
\extendedversiontrue

\usepackage{amssymb}
\usepackage{amsmath,amsfonts}
\usepackage{mathtools}
\usepackage{cuted}  
\usepackage{dsfont}
\usepackage{algpseudocode}
\usepackage{array}
\usepackage{booktabs}
\usepackage{textcomp}
\usepackage{stfloats}
\usepackage{url}
\usepackage{verbatim}
\usepackage{graphicx}
\usepackage[hidelinks]{hyperref}

\usepackage{subcaption}
\usepackage[affil-it]{authblk}
\usepackage{fancyhdr}
\usepackage{bm}
\usepackage[linesnumbered,ruled,vlined]{algorithm2e}
\SetKwInput{KwInput}{Input}
\SetKwInput{KwOutput}{Output}
\SetKw{KwAssert}{Assert}

\usepackage{amsthm}
\usepackage{setspace}

\newtheorem{lemma}{Lemma}
\newtheorem{theorem}{Theorem}
\newtheorem{proposition}{Proposition}
\newtheorem{corollary}{Corollary}[theorem]
\theoremstyle{definition}
\newtheorem{definition}{Definition}

\newtheorem{assumption}{Assumption}[]

\usepackage{setspace}

\usepackage{tikz}
\usetikzlibrary{shapes,arrows}
\usetikzlibrary{arrows,calc,positioning}
\tikzset{
	block/.style = {draw, rectangle,
		minimum height=1.2cm,
		minimum width=1.2cm},
	input/.style = {coordinate,node distance=1cm},
	output/.style = {coordinate,node distance=2cm},
	arrow/.style={draw, -latex,node distance=2cm},
	pinstyle/.style = {pin edge={latex-, black,node distance=1cm}},
	sum/.style = {draw, circle, node distance=1cm},
}
\definecolor{backgreen}{HTML}{E9F3DF}
\definecolor{backblue}{HTML}{DAE5EC}
\definecolor{backpurple}{HTML}{E5E0E8}
\definecolor{backorange}{HTML}{F2E9DF}
\definecolor{lightergray}{gray}{0.9}
\definecolor{lightgray}{gray}{0.85}

\usepackage[framemethod=TikZ]{mdframed}
\mdfdefinestyle{callout}{%
	linecolor=black,
	linewidth=1pt,%
	roundcorner=0pt,
	innertopmargin=4pt,
	innerbottommargin=4pt,
	innerrightmargin=5pt,
	innerleftmargin=5pt,
	leftmargin = 0pt,
	rightmargin = 0pt,
	backgroundcolor=lightergray
}

\include{macros.tex}

\include{macros_custom.tex}

\title{\LARGE \bf
	Evolutionary Analysis of Continuous-time Finite-state\\Mean Field Games with Discounted Payoffs}

\author{Leonardo Pedroso, Andrea Agazzi, W.P.M.H. (Maurice) Heemels, Mauro Salazar
	\thanks{This work was supported in part by the Eindhoven Artificial Intelligence Systems Institute (EAISI).}
	\thanks{L.~Pedroso, W.P.M.H.~Heemels and M.~Salazar are with the Control Systems Technology section, Department of Mechanical Engineering, Eindhoven University of Technology, The Netherlands (e-mail: \{l.pedroso,m.heemels,m.r.u.salazar\}@tue.nl). A.~Agazzi is with the Institute of Mathematical Statistics and Actuarial Science, Department of Mathematics and Statistics, University of Bern, Switzerland (e-mail: andrea.agazzi@unibe.ch).}%
}

\begin{document}
	
	\maketitle
	\pagestyle{plain} 
	\thispagestyle{plain}
	
	\begin{abstract}
		We consider a class of continuous-time dynamic games involving a large number of players. Each player selects actions from a finite set and evolves through a finite set of states. State transitions occur stochastically and depend on the player’s chosen action. A player’s single-stage reward depends on their state, action, and the population-wide distribution of states and actions, capturing aggregate effects such as congestion in traffic networks. 
		Each player seeks to maximize a discounted infinite-horizon reward. Existing evolutionary game-theoretic approaches introduce a model for the way individual players update their decisions in static environments without individual state dynamics. In contrast, this work develops an evolutionary framework for dynamic games with explicit state evolution, which is necessary to model many applications. We introduce a mean field approximation of the finite-population game and establish approximation guarantees. Since state-of-the-art solution concepts for dynamic games lack an evolutionary interpretation, we propose a new concept -- the Mixed Stationary Nash Equilibrium (MSNE) -- which admits one. We characterize an equivalence between MSNE and the rest points of the proposed mean field evolutionary model and we give conditions for the evolutionary stability of MSNE.
	\end{abstract}

\input{section/introduction.tex}
\input{section/model.tex}

\input{section/equilibria.tex}
\input{section/MSNE_rest.tex}
\input{section/evolutionary_stability.tex}

\input{section/MAC.tex}
\input{section/conclusion.tex}

\appendix
\ifextendedversion

\input{section/proofs_mfg_model.tex}

\fi
\input{section/proofs_MSNE.tex}
\ifextendedversion
\input{section/proofs_MSNE_rest.tex}
\input{section/evolutionary_stability_proofs.tex}

\fi

\bibliographystyle{IEEEtran}
\bibliography{../../../../Papers/_bib/references-gt.bib,../../../../Papers/_bib/references-c.bib,../../../../Publications/bibliography/parsed-minimal/bibliography.bib}

\end{document}

%% file: macros.tex
\usepackage{endnotes}

\newcommand{\dint}[0]{\mathrm{d}}

\DeclareMathOperator{\argmax}{\mathrm{argmax}}
\DeclareMathOperator{\EV}{\mathds{E}}

\DeclareMathOperator{\supp}{\mathrm{supp}}

\DeclareMathOperator{\col}{\mathrm{col}}

\DeclareMathOperator{\sign}{\mathrm{sgn}}

\DeclareMathAlphabet{\doublestruck}{U}{BOONDOX-ds}{m}{n}
\newcommand{\ones}[0]{\mathds{1}}
\newcommand{\zeros}[0]{\doublestruck{0}}

\newcommand{\N}[0]{\mathbb{N}}

\newcommand{\R}[0]{\mathbb{R}}

\newcommand{\Rnn}[0]{\mathbb{R}_{\geq0}}
\newcommand{\Rnp}[0]{\mathbb{R}_{\leq0}}

\newcommand{\Acal}[0]{\mathcal{A}}
\newcommand{\Bcal}[0]{\mathcal{B}}
\newcommand{\Ccal}[0]{\mathcal{C}}

\newcommand{\Ocal}[0]{\mathcal{O}}
\newcommand{\Pcal}[0]{\mathcal{P}}

\newcommand{\Scal}[0]{\mathcal{S}}

\newcommand{\Ucal}[0]{\mathcal{U}}

\newcommand{\Xcal}[0]{\mathcal{X}}

\usepackage{xstring}
\newcommand{\weblink}[1]{%
	\StrSubstitute{#1}{https://}{}[\strippedurl]%
	\href{#1}{{\small \color[RGB]{0,0,0} \url{\strippedurl}}}%
}

%% file: macros_custom.tex
\newcommand{\muSU}[0]{\mu}

\newcommand{\Rd}[0]{R_\mathrm{d}}

\newcommand{\Rdc}[0]{R_\mathrm{d}^c}
\newcommand{\Rrc}[0]{R_\mathrm{r}^c}

\newcommand{\UD}[0]{\Ucal_\mathrm{D}}
\newcommand{\UDc}[0]{\Ucal^c_\mathrm{D}}
\newcommand{\UDcbar}[0]{\bar{\Ucal}^c_\mathrm{D}}

\newcommand{\sE}[0]{\mathrm{E}}
\newcommand{\sAE}[0]{\mathrm{AE}}
\newcommand{\sAF}[0]{\mathrm{AF}}
\newcommand{\sF}[0]{\mathrm{F}}

\newcommand{\aN}[0]{0}
\newcommand{\aL}[0]{\mathrm{L}}
\newcommand{\aH}[0]{\mathrm{H}}

%% file: section/introduction.tex

\vspace{0.3cm}
\section{Introduction}

Systems composed of many interacting decision-makers often arise in economics, biology, engineering, and the social sciences. 
These settings share a common feature: individual agents interacting through population-level effects, that can be naturally modeled as games with many players. This paper studies a broad class of such models in which each player occupies a discrete state, takes actions at independent Poisson time instants, and receives an immediate reward that depends on its state, action, and the joint state–action distribution of the population \cite{GomesSaude2014,CarmonaDelarue2018}. Actions trigger stochastic state transitions, hence each decision affects future rewards, rendering the game dynamic. This distinguishes the model from static games, where players have no internal state and rewards depend only on the population's action distribution.

A central problem in game theory is to describe the outcomes of a game and how they emerge. Rules that characterize these outcomes are known as solution concepts, a celebrated example being the Nash equilibrium (NE). Evolutionary game theory \cite{SmithPrice1973} addresses this problem by defining revision protocols, which specify how players adjust their decisions through simple, myopic rules. In static games, these protocols induce revision dynamics on the population’s action distribution, and the rest points of these dynamics provide natural solution concepts with an evolutionary interpretation, which often coincide with NE \cite{Sandholm2010}.

\begin{mdframed}[style= callout]
However, no analogous evolutionary interpretation exists for established solution concepts in dynamic games. Prior work overwhelmingly relies on so-called behavioral equilibria, in which all players adopt the same mapping from states to action distributions \cite{AdlakhaJohariEtAl2015,WiecekAltman2015,Wiecek2020}. Although analytically convenient, this assumption forces homogeneity across players and prevents meaningful modeling of individual revisions. Consequently, existing solution concepts for dynamic games lack an evolutionary interpretation.
\end{mdframed}

Some partial progress has been made for dynamic games \cite{AltmanHayel2010,FleschParthasarathyEtAl2013,BrunettiHayelEtAl2018}. However, these works address only a specific setting in which interactions occur exclusively in pairs, and none rely on an individual-based evolutionary model analogous to the evolutionary game theory for static games. For further discussion of the limitations of these approaches, see \cite[Section~I]{PedrosoAgazziEtAl2025MFGAvg}.
Recently, \cite{PedrosoAgazziEtAl2025MFGAvg,PedrosoAgazziEtAl2025MFGAvgII} initiated a formal evolutionary analysis of continuous-time finite-state mean field dynamic games with average payoffs. In this paper, we extend that analysis to the discounted-payoff setting, which presents a fundamental difference from the average-payoff case: \emph{the payoff of a state–action map depends on the initial state} from which it is used. This dependence introduces asymmetries in the evolutionary dynamics and leads to qualitatively different evolutionary phenomena.

The main contributions of this paper are fourfold.
First, we introduce a novel solution concept for the discounted-payoff setting, which we call the mixed stationary Nash equilibrium (MSNE), and establish its existence under mild conditions.
Second, we formulate a mean field evolutionary model for this class of dynamic games and show that its solutions approximate those of the $N$-player game as $N \to \infty$. Third, we characterize the relationship between MSNE and evolutionary rest points, proving that they coincide under broad families of revision protocols. Fourth, we conduct a preliminary analysis of the evolutionary stability of MSNE.

\subsection{Notation}	

For $N\in \N$, the set of consecutive positive integer numbers $\{1,2,\ldots, N\}$ is denoted by $[N]$.
The $i$th entry of a vector $x\in \R^n$ is denoted by $x_i$. The Euclidean norm of a vector $x\in \R^n$ is denoted by $||x||$. The $n$ dimensional vector of zeros and ones are denoted by $\zeros_{n}$ and $\ones_n$, respectively. When clear from context, the subscript $n$ will be dropped to streamline notation. The sign of $x\in \R$ is denoted by $\sign(x)$ and takes the values of $-1$, $0$, or $1$ if $x<0$, $x=0$, or $x>0$, respectively. The column-wise concatenation of a finite number of vectors $x^1, x^2, \ldots, x^K$ is denoted by $\col(x^1, x^2, \ldots, x^K)$. The indicator function of $a\in \Xcal$ is denoted by $\delta_a:\Xcal \to \{0,1\}$ and $\delta_a(x) = 0$ if $x\neq a$ and $\delta_a(x) = 1$ if $x = a$. The support of a function $f:\Xcal\to \R$ is denoted by $\supp(f):=\{x\in \Xcal: f(x)\neq 0\}$. 
The interior of a set $\Acal$ is denoted by $\mathrm{int}(\Acal)$. Given sets $\Xcal_1,\Xcal_2,  \ldots,  \Xcal_K$,  the Cartesian product $\Xcal_1 \times \Xcal_2 \times \cdots \times \Xcal_K$ is denoted by $\bigtimes_{k= 1}^{K}\Xcal_k$. The expected value of a random variable (r.v.) $Z$ is denoted by $\EV[Z]$. The set of all Borel probability measures on $\Acal$ is denoted by $\Pcal(\Acal)$. Given a probability measure $\eta  \in \Pcal(\Acal)$, the mass on $a\in \Acal$ is denoted by $\eta(a)$.
In this paper, to characterize the distribution of mass of a population of mass $m>0$ over elements of a finite set $\Acal$ we use vectors $\mu \in X_{\Acal}:= \{ \nu \in  \Rnn^{|\Acal|} : \ones^\top \nu = m\}$. For the sake of clarity, by abuse of notation, the mass on $a\in \Acal$ is denoted by $\mu[a]$ and the mass on a subset $\Bcal\subseteq \Acal$ is denoted by $\mu[\Bcal]:= \sum_{a\in \Bcal}\mu[a]$.

%% file: section/model.tex
\section{Model}

In this section, we describe the model for a population of $N$ players and the mean field model approximation as $N \to \infty$. This model is very similar to the one in \cite{PedrosoAgazziEtAl2025MFGAvg} for average payoffs and it is presented in what follows for the sake of completeness.

\subsection{Finite Population Model}

The finite-population model is described by:
\begin{itemize}
	
	\item \emph{Population}: There are $N \in \N$ players which are spread across $C\in \N$ classes (also called subpopulations) with similar needs. We denote the class of a player $i\in [N]$ by $c^i$, which is time-invariant. The set of players that are in a class $c\in [C]$ is denoted by $\Ccal_c :=\{i\in [N]: c^i= c\}$.  The mass of players in a class $c\in \Ccal$ is denoted by $m^c := |\Ccal_c|/N$.
	
	\item \emph{Time}: Each player makes decisions in continuous time. Each player $i\in \Ccal_c$ is equipped with a Poisson clock with rate $\Rdc >0$ (which is equal to the rate of all other players in the same class). Each time the clock of a player rings, they take an action. We assume that clocks of different players are independent. The time of the $k$-th clock ring of a player $i\in [N]$ is characterized by a random variable (r.v.) $t^{i}_k$.
	
	\item \emph{States}: At each time $t \in [0,\infty)$, each player $i\in \Ccal_c$ has an individual state from a finite set of states $\Scal^c$. The player's state evolves stochastically with their decisions and is characterized by a r.v.\ $s^{i}(t)$. As a result, a realization of $s^{i}(t)$ has a piecewise-constant time evolution with discontinuities when the clock of the $i$-th player rings. We also define $p^c := |\Scal^c|$. 
	
	\item \emph{Actions}: The actions available to a player $i\in \Ccal_c$ in state $s\in \Scal^c$ are in the nonempty finite set $\Acal^c(s)$. We denote by $\Acal^c := \bigcup_{s\in \Scal} \Acal^c(s)$ the set of all actions available to a player of class $c$. The action that a player $i\in [N]$ would take at time $t$ if their clock were to ring is characterized by a r.v.\ $a^{i}(t)$. We also define $q^c := |\Acal^c|$. 
	
	\item \emph{State transitions}: Upon an action of a player, their state evolves according to a Markov transition kernel $\phi^c: \Scal^c \times \Acal^c \to \Pcal(\Scal^c)$. We denote the distribution of the new state of a player in state $s \in \Scal^c$ that takes action $a\in \Acal^c(s)$ by $\phi^c(\cdot | s,a)$.
	
	\item \emph{State-action distribution}: The empirical joint state-action distribution of class $c\in [C]$ at time $t$ is characterized by a r.v.\ $\hat{\mu}^c_{\Scal \times \Acal}(t)$ with support in $X^c_{\Scal \times \Acal}:=\{\nu \in \Rnn^{p^cq^c} : \ones^\top\nu = m^c\}$. Recall that, by abuse of notation, $\hat{\mu}^c_{\Scal \times \Acal}[s,a](t)$ is the r.v.\ associated with the mass on $s\in \Scal^c$ and $a\in \Acal^c$ and it is given by $\hat{\mu}^c_{\Scal \times \Acal}[s,a](t) := \frac{1}{N}\sum_{i\in \Ccal_c} \delta_{s^{i}(t)}(s) \delta_{a^{i}(t)}(a)$. The concatenation of the empirical joint state-action distributions for all classes is denoted by $\hat{\mu}_{\Scal \times \Acal} = \col(\hat{\mu}^c_{\Scal \times \Acal}, c\in [C])$ with support in $X_{\Scal \times \Acal} := \bigtimes_{c\in [C]}X^c_{\Scal \times \Acal}$.
	
	\item \emph{Single-stage reward}: The single-stage reward of a player $i\in \Ccal_c$ is modeled by a real-valued function $r^c: \Scal^c \times \Acal^c \times X_{\Scal \times \Acal} \to \R$.  Specifically, the single-stage reward of a player in state $s\in \Scal^c$ that takes action $a\in \Acal^c(s)$ at time $t$ is $r^c(s,a,\hat{\mu}_{\Scal\times\Acal}(t))$. Notice that $N\sum_{c\in [C]: a\in \Acal^c}\Rdc\hat{\mu}^c_{\Scal\times\Acal}[\Scal^c,a](t)$ corresponds to the expected flow of players taking action $a$, which can, for instance, be used to model a decreasing reward upon congestion of a resource. 
	
	\item \emph{Payoff}: The payoff of a player $i\in [N]$ is the discounted infinite-horizon reward which is given by
	\begin{equation*}
		J^i := \EV\left[\,\sum_{k=1}^\infty \beta^k r^{c^i}\left(s^{i}(t^i_k),a^{i}(t^i_k),\hat{\mu}_{\Scal \times \Acal}(t^i_k)\right)\right],
	\end{equation*} 
	where  $\beta \in (0,1)$ is the discount factor. 
\end{itemize}

\subsection{Information Structure}\label{sec:policies}

The information structure specifies what each player observes when making decisions. Given this information, a player selects one action. The mapping from the available information to the chosen action is referred to as a \emph{policy}. In our framework, players possess only limited information, summarized as follows:

\begin{itemize}
	\item Players cannot observe the population-level distribution of states or actions.
	\item Players do not retain a history of their past states or actions. When their personal clock rings, they only observe their current state.
	\item Players do not anticipate future changes in the population’s state-action distribution. The payoff of any policy perceived by a player is evaluated as if the aggregate state-action distribution were stationary and the policy were used perpetually.
\end{itemize}
Policies compatible with this information structure are, respectively, called \emph{oblivious}, \emph{Markov}, and \emph{stationary}. Formally, for each class $c\in[C]$, such a policy is a map $u:\Scal^c \to \Pcal(\Acal^c)$ assigning to each state a probability distribution over feasible actions. The set of admissible policies for class $c$ is
\begin{equation*}
	\Ucal^c := \left \{ u:\Scal^c \to \Pcal(\Acal^c)  \;| \; \supp(u(s)) \subseteq \Acal^c(s) \; \forall s\in \Scal^c \right\}.
\end{equation*} 
In general, policies in $\Ucal^c$ may \emph{randomize} over actions. A policy is  \emph{deterministic} if each state is mapped to a single action with probability one. This deterministic subset is denoted by $\UDc \subset \Ucal^c$ and is formally defined as
\begin{equation*}
	\UDc := \left \{ u \in \Ucal^c \; |\; \forall s\in \Scal^c\; \exists a\in \Acal^c: \supp(u(s)) = \{a\}\right\}\!.
\end{equation*}
We also define $n^c := |\UDc|$.
Whenever the meaning is unambiguous, we write deterministic policies using the shorthand $u(s) = a$, instead of the probability measure $u(s) = \delta_a(\cdot)$. Each player  $i\in \Ccal_c$ is assumed to use a deterministic policy that is characterized by the r.v.\ $u^i(t)$ at time $t$. Section~\ref{sec:ev_dynamics} introduces a model describing how these policies evolve over time. The empirical joint state-policy distribution of class $c\in [C]$ is characterized by a r.v.\ $\hat{\mu}^c(t)$ with support in $X^c :=\{\nu \in \Rnn^{p^cn^c} : \ones^\top\nu = m^c\}$, which, by abuse of notation, is given by $\hat{\mu}^{c}[s,u](t) := \frac{1}{N}\sum_{i\in \Ccal_c}^{N} \delta_{s^{i}(t)}(s) \delta_{u^{i}(t)}(u)$. The concatenation of the empirical joint state-policy distributions for all classes is denoted by $\hat{\mu} = \col(\hat{\mu}^c, c\in [C])$ with support in $X:= \bigtimes_{c\in [C]}X^c$.

For each class $c\in [C]$, we write the discounted infinite-horizon reward of a policy $u\in \UDc$ starting from an initial state distribution $\eta^c_0 \in \Pcal(\Scal^c)$ when the aggregate state-action distribution $\mu_{\Scal \times \Acal} \in X_{\Scal \times \Acal}$ is constant as
\begin{equation}\label{eq:def_J}
	J^c(u,\eta^c_0,\mu_{\Scal\times\Acal}) :=  \EV\left[\,\sum_{k=1}^\infty \beta^kr(s_k,u(s_k),\mu_{\Scal \times \Acal})\right],
\end{equation}
where $s_k \sim  \eta^c_k$ and $\eta^c_{k}\in \Pcal(\Scal^c)$ is characterized by $\eta^c_{k}(s) = \sum_{s^\prime \in \Scal^c}\sum_{a^\prime \in \Acal^c}\phi^c(s|s^\prime, a^\prime)u(a^\prime|s^\prime)\eta^c_{k-1}(s^\prime)$ for all $s\in \Scal^c$. By abuse of notation, whenever $\eta^c_0$ has all the mass at a state $s_0\in \Scal^c$, we, alternatively, write $J^c(u,s_0,\mu_{\Scal\times\Acal})$.

\begin{mdframed}[style=callout]
	In contrast to the average-payoff setting analyzed in \cite{PedrosoAgazziEtAl2025MFGAvg,PedrosoAgazziEtAl2025MFGAvgII}, the \emph{payoff of a policy depends on the initial state} from which it is used.
\end{mdframed}

\subsection{Evolutionary Decision Model}\label{sec:ev_dynamics}

Evolutionary approaches, which are classically studied in static settings \cite{Sandholm2010}, provide the basis for the dynamic evolutionary model we adopt. In contrast to coordination-based population adjustments (which are unnatural in a large population), players in our model revise their policies individually, capturing realistic features such as inertia and myopic behavior. By abuse of notation, for a class $c\in [C]$ and a state $s\in \Scal^c$, we denote the vector of the mass on each policy as $\hat{\mu}^c[s,\cdot](t) := \col(\hat{\mu}^c[s,u](t),u\in \UDc) \in \Delta^c_{\UD}$,  where $\Delta^c_{\UD} := \{\sigma \in \Rnn^{n^c}: \ones^\top \sigma \leq m^c\}$. Henceforth, the time dependence is oftentimes dropped for conciseness. The evolutionary model is described by: 

\begin{itemize}
	\item \emph{Time}: Each player $i\in \Ccal_c$ has a Poisson revision clock with rate $\Rrc >0$ (which is equal to the rate of all other players in the same class). Each time the clock of a player rings, they have the opportunity to revise the policy that they are currently using. We assume that action and revision clocks of all players are independent.
	
	\item \emph{Policy transitions}: Upon a revision opportunity of a player, their policy choice evolves according to a \emph{revision protocol}. A revision protocol of a class $c\in [C]$ is a map $\rho^c : \R^{n^c} \times \Delta^c_{\UD} \to \Rnn^{n^c\times n^c}$. The component associated with the pair $(u,v) \in \UDc\times \UDc$ is denoted, by abuse of notation, by $\rho^c_{uv}$. Specifically, a player in state $s\in \Scal^c$ using policy $u\in \UDc$ switches to policy $v\in \UDc$ with a switch rate $\rho^c_{uv}(F^{c,s}(\hat{\mu}),\hat{\mu}^c[s,\cdot])$. Here, $F^{c,s}(\hat{\mu})$ is a vector of the discounted infinite-horizon payoffs for each policy in $\UDc$ starting from state $s$, i.e., 
	\begin{equation}\label{eq:def_Fs}
		F^{c,s}(\hat{\mu}) := \col\left(J^c(u,s,\hat{\mu}_{\Scal\times \Acal}), u\in \UDc\right),
	\end{equation}
	where $\hat{\mu}_{\Scal\times \Acal}$ is written as a function of $\hat{\mu}$ as $\hat{\mu}^c_{\Scal\times\Acal}[s,a] = \sum_{u\in \UD}\hat{\mu}^c[s,u]u(a|s)$ for all $c\in [C]$, all $s\in \Scal^c$, and all $a\in \Acal^c$ and the policy ordering is consistent with the definition of $\hat{\mu}^c[s,\cdot]$. For the sake of clarity, by abuse of notation, we denote the component associated with policy $u\in\UDc$ by $F^{c,s}_u(\mu)$.
\end{itemize}

Intuitively, if a player $i\in \Ccal_c$ in state $s\in \Scal^c$ using policy $u\in \UDc$ receives a revision opportunity, they switch to a policy $v\in \UDc$ with probability $\rho^c_{uv}(F^{c,s}(\hat{\mu}),\hat{\mu}^c[s,\cdot])/\Rrc$, and they continue to use the same policy with probability $1-\sum_{v\neq u}\rho^c_{uv}(F^{c,s}(\hat{\mu}),\hat{\mu}^c[s,\cdot])/\Rrc$.
The literature on evolutionary game theory identifies three physically meaningful classes of revision protocols \cite[Chap.~5]{Sandholm2010}, which are slightly redefined for our setting as follows.

\begin{definition}[Imitative]\label{def:imitative}
	Consider a revision protocol $\rho^c$ that is defined as $\rho^c_{uv}(F^c,\sigma) = r^c_{uv}(F^c,\sigma)\sigma_v/(\ones^\top\sigma)$ if $\sigma \neq \zeros$ and is null otherwise, where $r^c :\R^{n^c} \times \Delta_{\UD}^c\to \Rnn^{n^c\times n^c}$ is a Lipschitz continuous conditional imitation rate map  with monotone net conditional imitation rates, i.e., $F^c_v \geq F^c_u \iff r^c_{kv}(F^c,\sigma) -   r^c_{vk}(F^c,\sigma)  \geq  r^c_{ku}(F^c,\sigma) -   r^c_{uk}(F^c,\sigma), \forall F^c \in \R^{n^c}\;\forall \sigma \in  \Delta_{\UD}^c\;\forall u,v,k \in \UDc$. Then $\rho^c$ is said to be an \emph{imitative} revision protocol. 
\end{definition}

\begin{definition}[Excess payoff]\label{def:excess}
	Consider a revision protocol $\rho^c$ defined as $\rho^c_{uv}(F^c,\sigma) = \tau_{v}^c(\hat{F}^c)$,  where $\hat{F}^c$ is the excess payoff vector that is defined by  $\hat{F}^c := F^c - \ones {F^c}^\top\sigma/(\ones^\top \sigma)$ if $\sigma \neq \zeros$ and is the vector of zeros otherwise,  and $\tau^c :\R^{n^c} \to \Rnn^{n^c}$ is a Lipschitz continuous rate map that satisfies acuteness, i.e.,
	$\hat{F}^c \in \R^{n^c} \setminus \Rnp^{n^c} \!\implies \! \tau^c(\hat{F}^c)^\top\hat{F}^c > 0$. Then $\rho^c$ is called an \emph{excess payoff} revision protocol. Furthermore, $\rho^c$ is said to be a \emph{separable excess payoff} revision protocol if $\tau_{v}^c(\hat{F}^c) \equiv \tau_{v}^c(\hat{F}_v^c)$. 
\end{definition}

\begin{definition}[Pairwise comparison]\label{def:pairwise_cmp}
	Consider a revision protocol $\rho^c$ defined as $\rho^c_{uv}(F^c,\sigma) = \tau^c_{uv}(F^c)$, where $\tau :\R^{n^c} \to \Rnn^{n^c}$ is a Lipschitz continuous rate map  that is sign-preserving, i.e., $\sign(\tau^c_{uv}(F)) = \sign(\max(0,F^c_v-F^c_u)), \forall F^c \in \R^{n^c} \; \forall u,v \in \UDc$. Then $\rho^c$  is called a \emph{pairwise comparison} revision protocol.
\end{definition}

For a more detailed description of these families and of the meaningfulness of the evolutionary dynamics generated by them refer to \cite[Part~II]{Sandholm2010}. 

\begin{mdframed}[style=callout]
	In a discounted-payoff setting, since the payoff perceived by a player depends on their state, the revision of their policy relies on comparing its payoff to the payoff of other policies evaluated from that same state. This difference is subtle but prevents the direct use of the analysis results of the average-payoff setting in \cite{PedrosoAgazziEtAl2025MFGAvg,PedrosoAgazziEtAl2025MFGAvgII}. Indeed, the asymmetries in the revision flows between two policies from different states will lead to qualitatively different evolutionary phenomena.
\end{mdframed}

\subsection{Assumptions}\label{sec:model_ass}

In what follows, we introduce mild regularity conditions on the model. First, we assume global Lipschitz continuity of the single-stage reward, which is reasonable for physically meaningful rewards such as congestion models. This is required for the existence of equilibria and existence and uniqueness of solutions to the equations that will describe the evolutionary dynamics. Second, we make an assumption to ensure that the revision switching probabilities are well defined.

\begin{assumption}\label{ass:cont}
	For all $c \in [C]$, all $s\in \Scal^c$, and all $a\in \Acal^c$ the single-stage reward function $r^c(s,a,\mu_{\Scal\times \Acal})$ is Lipschitz continuous w.r.t.\ $\mu_{\Scal\times \Acal}$ in $X_{\Scal \times \Acal}$ w.r.t\ the Euclidean norm. 
\end{assumption}

\begin{assumption}\label{ass:rev_protocol}
	For all $c\in [C]$, $\rho^c$ satisfies
	\begin{equation*}
		1 - \sup\nolimits_{\mu \in X} \sum\nolimits_{v\in \UDc \setminus \{u\}} \rho^c_{uv}(F^{c,s}(\mu),\mu^c[s,\cdot])/\Rrc \;\geq 0.
	\end{equation*}
\end{assumption}

\vspace{0.3cm}

\section{Mean Field Approximation}

Considering a continuum of players instead of a finite number allows to describe the revision dynamics by the evolution of the joint state-policy distribution of the population. The joint state-policy distribution at time $t$ is denoted by  $\mu(t) \in X$. Henceforth, the time dependence is oftentimes dropped for conciseness. 

Intuitively, in an infinitesimal interval of time $\dint t$, for a class $c$, the difference in the mass in state $s\in \Scal^c$ evolves according to the Markov kernel $\phi^c$ and the difference in the mass in policy $u\in \UDc$: (i)~increases by the proportion of revision clock rings in other policies that switch to policy $u$; and (ii)~decreases by the proportion of revision clock rings in policy $u$ that switch to another policy, i.e., $\forall s\in \Scal^c \; \forall u\in \UDc$
\begin{equation*}
	\begin{split}
		\dint \mu^c[s,u] &=   \sum_{s^\prime\in \Scal^c}  \sum_{a^\prime\in \Acal^c}\Rdc\mu^c[s^\prime,u] \dint t\phi^c(s|s^\prime,a^\prime)u(a^\prime|s^\prime)\\
		&- \Rdc\mu^c[s,u]\dint t \sum_{s^\prime\in \Scal^c} \sum_{a\in \Acal^c}\phi^c(s^\prime|s,a)u(a|s),\\
		&+ \sum_{u^\prime \in \UDc} \Rrc\mu^c[s,u^\prime] \dint t \rho^c_{u^\prime u}(F^{c,s}(\mu),\mu^c[s,\cdot])/\Rrc \\
		&- \Rrc \mu^c[s,u] \dint t\sum_{u^\prime \in \UDc} \rho^c_{uu^\prime}(F^{c,s}(\mu),\mu^c[s,\cdot])/\Rrc.
	\end{split}
\end{equation*}
When $\dint t \to 0$ this balance equation can be written as 
\begin{equation}\label{eq:ODE_mu_ev}
	\dot{\mu}^c[s,u] =f_{s,u}^{c,\mathrm{d}}(\mu) + 	f_{s,u}^{c,\mathrm{r}}(\mu), 
\end{equation}
where
\begin{equation}\label{eq:fd_fr}
	\begin{split}
	f_{s,u}^{c,\mathrm{d}}(\mu) =&  \Rdc \sum_{s^\prime \in \Scal^c} \sum_{a^\prime \in \Acal^c}\!\!\!\phi^c(s|s^\prime,a^\prime)u(a^\prime|s^\prime)\mu^c[s^\prime,u]  \\
	-& \Rdc\mu^c[s,u] \\
	f_{s,u}^{c,\mathrm{r}}(\mu) =& \sum_{u^\prime \in \UDc}  \mu^c[s,u^\prime] \rho^c_{u^\prime u}(F^{c,s}(\mu),\mu^c[s,\cdot])  \\
	- &\mu^c[s,u] \sum_{u^\prime \in \UDc}   \rho^c_{uu^\prime}(F^{c,s}(\mu),\mu^c[s,\cdot]).
\end{split}
\end{equation}
The ordinary differential equation (ODE) in \eqref{eq:ODE_mu_ev} is called the \emph{mean dynamic} or \emph{master equation}. Due to the aforementioned regularity assumptions, the mean dynamic is well defined and it approximates the finite-population model well, as formally detailed in the following result.

\begin{theorem}\label{th:approx_ODE_mu_ev}
		Consider an imitative, excess payoff, or pairwise comparison revision protocol $\rho^c$ for each class $c\in [C]$. Under Assumptions~\ref{ass:cont}-\ref{ass:rev_protocol}, a solution to the master equation, characterized by \eqref{eq:ODE_mu_ev}, with initial condition $\muSU(0)\in X$ exists in $t\in [0,\infty)$, is unique, and is Lipschitz continuous w.r.t.\ $\muSU(0)$. Moreover, if $\lim_{N\to \infty}\hat{\mu}(0)= \mu(0)$ almost surely, then for all $T<\infty$ $\hat{\mu}(t)$ converges in probability to $\mu(t)$ for all $t\in [0,T]$ as $N\to \infty$.
\end{theorem}
\begin{proof}
	\ifextendedversion
		See Appendix~\ref{sec:proof_th_approx_ODE_mu_ev}.
	\fi
\end{proof}

%% file: section/equilibria.tex
\section{Solution Concept}\label{sec:equilibria}

We now introduce a solution concept for the mean field game model under study. As in any strategic setting, the value of a solution concept lies in its ability to capture the long-run outcome of the interactions between players in the population. Crucially, there are two conceptually different ways to interpret the population’s policy profile: a  \emph{behavioral}  interpretation and a \emph{mixed} interpretation.

The prevailing literature on continuous-time, finite-state stochastic dynamic games with many players (and related mean field game formulations) almost exclusively adopts the behavioral stationary Nash equilibrium (BSNE) as the benchmark solution concept; see, for example, \cite{AdlakhaJohariEtAl2015, WiecekAltman2015, Wiecek2020}. Informally:

\begin{mdframed}[style=callout]
	A \emph{behavioral stationary Nash equilibrium} (BSNE) is an equilibrium condition whereby all players of the same class $c\in [C]$ use the same (randomized) policy $u_c \in \Ucal^c$ (the population uses a behavioral policy) such that: (i)~the resulting state distribution is stationary; and (ii)~no player can unilaterally deviate from $u_c$ to another policy $v\in \Ucal^c$ to increase their payoff. 
\end{mdframed}

While analytically convenient, this assumption that every player in a class must adopt an identical randomized policy is not motivated by any physically meaningful principle. In many settings, such as ours, it is more natural to allow players within a class to differ in the policies they use. This leads us to an alternative equilibrium notion that incorporates such \emph{heterogeneity}: the mixed stationary Nash equilibrium (MSNE). Informally:

\begin{mdframed}[style=callout]
	A \emph{mixed stationary Nash equilibrium} (MSNE) is an equilibrium condition whereby each player of a class $c\in[C]$ uses a deterministic policy $u \in \UDc$ (the population uses a mixed policy) such that: (i)~the resulting state distribution is stationary; and (ii)~no player can unilaterally deviate from $u$ to another policy $v \in \UDc$ to increase their payoff. 
\end{mdframed}

The intuitive definition of behavioral and mixed equilibria is identical to the framework with average payoffs in \cite{PedrosoAgazziEtAl2025MFGAvg}. Refer to \cite[Sec.~III]{PedrosoAgazziEtAl2025MFGAvg} for a more detailed discussion.

\subsection{Definition of BSNE and MSNE}

In this section, we present the formal definitions of BSNE and MSNE. 

\begin{definition}[BSNE]\label{def:BSNE}
	For each class $c\in [C]$, consider a policy $u_c\in \Ucal^c$ and a state distribution $\eta^c \in \Pcal(\Scal^c)$. The collection $(u_c,\eta^c)_{c\in[C]}$ is said to be a BSNE in the discounted payoff mean field game if 
	\begin{equation*}
		J^c(u_c,\eta^c,\mu_{\Scal\times\Acal})  \geq J^c(v,\eta^c,\mu_{\Scal\times\Acal})\quad \forall c\in[C]\;\forall v\in \Ucal^c,
	\end{equation*}
	where $\mu_{\Scal\times\Acal} \in X_{\Scal \times \Acal}$ is characterized by $\mu_{\Scal\times\Acal}^c[s,a] = m^c\eta^c(s)u(a|s)\; \forall c\in [C]\;\forall s\in \Scal^c \; \forall a\in \Acal^c$  and 
	\begin{equation*}
		\eta^c(s) \!=\!\! \!\sum_{s^\prime \!\in \Scal^c} \!\sum_{a^\prime\! \in \Acal^c} \!\!\phi^c(s|s^\prime,a^\prime)u(a^\prime|s^\prime)\eta^c(s^\prime) ,\: \forall c\!\in\! [C]\, \forall s \!\in\! \Scal^c\!\!.\!
	\end{equation*}
\end{definition}

\begin{definition}[MSNE]\label{def:MSNE}
	A joint state-policy distribution $\mu \in X$ is said to be a MSNE in the discounted payoff mean field game if for all $c\in [C]$, all $s\in \Scal^c$, and all $u\in \UDc$
	\begin{equation}\label{eq:MSNE_u}
\mu^c[s,u] >0 \implies   F^{c,s}_u(\mu)  \geq F^{c,s}_v(\mu) \; \forall  v\in \UDc
	\end{equation}
	and for all $c\in [C]$, all $s\in \Scal^c$, and all $u\in \UDc$
	\begin{equation}\label{eq:MSNE_s}
		\mu^c[s,u] = \sum_{s^\prime \in \Scal^c} \sum_{a^\prime \in \Acal^c}\phi^c(s|s^\prime,a^\prime)u(a^\prime|s^\prime)\mu^c[s^\prime,u].
	\end{equation}
\end{definition}

It is interesting to note the particularly intuitive definition for the MSNE. It is an equilibrium condition whereby every player follows a deterministic policy in steady-state and has no incentive to switch from their policy to any other deterministic policy from any state that they visit. This intuitive definition is instrumental to study the relation with rest points of the evolutionary dynamics in Section~\ref{sec:rest_point_MSNE} and study the evolutionary stability of MSNE in Section~\ref{sec:stability}.

\subsection{Existence}

In this section, we establish the existence of at least one MSNE. We resort to writing the MSNE condition as a fixed point of a set-valued map and then using Kakutani's fixed point theorem to establish the existence of a fixed point. The argument presented here is fundamentally different from that used for the analogous average-payoff result in \cite[Theorem~1]{PedrosoAgazziEtAl2025MFGAvg}.

\begin{theorem}\label{th:existence_MSNE}
	Under Assumption~\ref{ass:cont}, there exists at least one MSNE.
\end{theorem}
\begin{proof}
	\ifextendedversion
	See Appendix~\ref{sec:proof_th_existence_MSNE}.%
	\fi
\end{proof}

\begin{mdframed}[style=callout]
	A well-known result for finite Markov decision processes is that discounted-payoff-optimal policies are average-payoff-optimal for every discount factor close enough to one \cite{Blackwell1962}. This result can be expected to show that discounted-payoff MSNE are also average-payoff MSNE for every discount factor close enough to one.
\end{mdframed}

%% file: section/MSNE_rest.tex

\section{MSNE and Evolutionary Equilibria}\label{sec:rest_point_MSNE}

In this section, we study the relation between a rest point of the evolutionary dynamics \eqref{eq:ODE_mu_ev} and the MSNE solution concept. The first result in that regard is that every MSNE is a rest point of the evolutionary dynamics for all classes of revision protocols defined in Section~\ref{sec:ev_dynamics}.

\begin{theorem}\label{th:MSNE_is_rest}
	Consider an imitative, excess payoff, or pairwise comparison revision protocol $\rho^c$ for each class $c\in [C]$. If $\mu \in X$ is a MSNE, then $\mu$ is a rest point of the evolutionary dynamics \eqref{eq:ODE_mu_ev}.
\end{theorem}
\begin{proof}
	\ifextendedversion
		See Appendix~\ref{sec:proof_th_MSNE_is_rest}.
	\fi
\end{proof}

From Theorem~\ref{th:MSNE_is_rest} it follows that every MSNE is a rest point of the evolutionary dynamics for imitative, excess payoff, and pairwise comparison revision protocols.  However, for the converse to be true, stronger conditions are required, which only hold for pairwise comparison revision protocols, as shown in the following result.

\begin{theorem}\label{th:rest_is_MSNE}
	Consider a pairwise comparison revision protocol $\rho^c$ for each class $c\in [C]$.  If $\mu \in X$ is a rest point of the evolutionary dynamics \eqref{eq:ODE_mu_ev}, then $\mu$ is a MSNE.
\end{theorem}
\begin{proof}
	\ifextendedversion
		See Appendix~\ref{sec:proof_rest_is_MSNE}.
	\fi
\end{proof}


Since imitative revision protocols rely on revisions that imitate the policies of other players, if a policy $u \in \UDc$ of a class $c\in [C]$ does not have any mass in the initial condition, i.e., $\sum_{s\in \Scal^c}\mu^c[s,u](0) = 0$, then  $\sum_{s\in \Scal^c}\mu^c[s,u](t) = 0$ for all $t\geq 0$. This is the only reason why a rest point of an imitative revision protocol is not necessarily a MSNE. The same observation and qualitative behavior applies to the average-payoff setting. However, any perturbation that places a small mass on payoff maximizing policies renders those non-MSNE rest points unstable. This stability analysis is carried out formally in Section~\ref{sec:unst_nonMSNE}.

\begin{mdframed}[style = callout]
	Remarkably, in static games \cite[Theorem~5.5.2]{Sandholm2010} and dynamic games with average payoffs \cite[Theorems~4 and~5]{PedrosoAgazziEtAl2025MFGAvg} there is an equivalence between MSNE and the rest points of the evolutionary dynamics not only for pairwise comparison revision protocols but also for excess payoff revision protocols. In a setting with discounted payoffs, since the payoff of a given policy depends on the initial state of the player, uneven revision flows are introduced in excess payoffs dynamics, which may lead to rest points that are not MSNE. A numerical example of such behavior is available in an open-access repository at \weblink{https://github.com/fish-tue/evolutionary-mfg-discounted}.
\end{mdframed}

%% file: section/evolutionary_stability.tex

\section{Evolutionary Stability of MSNE}\label{sec:stability}

In this section, we turn to the local stability of MSNE. For the sake of simplicity, we assume the presence of regularization noise in the state dynamics such that every state is reachable from any other state under any deterministic policy. This assumption, motivated by the intrinsically stochastic nature of real-world decision dynamics, plays a key role in our analysis by preventing degenerate cases. It is formally described below and guarantees the irreducibility of the state transition kernel induced by each policy. For a detailed overview of the standard Markov chain analysis tools used in this paper we refer to \cite{Norris1997}.

\begin{assumption}\label{ass:noise}
	For all $c\in [C]$, and all $u\in \UDc$, the Markov kernel $\phi^{c,u} : \Scal^c \to \Pcal(\Scal^c)$ defined by $\phi^{c,u}(s|s^\prime) =  \sum_{a^\prime \in \Acal^c}\phi^c(s|s^\prime,a^\prime)u(a^\prime|s^\prime)$, is irreducible and, thus, the continuous-time Markov process generated by $Q^{c,u} = \Rdc(\phi^{c,u}-I)$ admits a unique stationary state distribution that has full support.
\end{assumption}

\subsection{Instability of non-MSNE rest points}\label{sec:unst_nonMSNE}

Recall that in the analysis in Section~\ref{sec:rest_point_MSNE}, for imitative revision protocols, a rest point of the evolutionary dynamics is not necessarily a MSNE. The following result shows that non-MSNE rest points are unstable under the evolutionary dynamics \eqref{eq:ODE_mu_ev}.

\begin{theorem}\label{th:imitative_rest_unstable}
	Under Assumptions~\ref{ass:cont}--\ref{ass:noise}, consider an imitative or pairwise comparison revision protocol $\rho^c$ for each class $c\in[C]$. Let $\mu^\star$ be a rest point of the evolutionary dynamics \eqref{eq:ODE_mu_ev}. If $\mu^\star$ is not a MSNE, then $\mu^\star$ is not Lyapunov stable under \eqref{eq:ODE_mu_ev} and no solution trajectory of \eqref{eq:ODE_mu_ev} with $\mu(0)\in \mathrm{int}(X)$ converges to $\mu^\star$.
\end{theorem}
\begin{proof}
		\ifextendedversion
	See Appendix~\ref{sec:proof_imitative_rest_unstable}.
	\fi
\end{proof}

The following corollary is the reciprocal of Theorem~\ref{th:imitative_rest_unstable} and allows to conclude that if a trajectory with a non-degenerate initial condition converges to a rest point, the rest point is a MSNE. As a result, under a very weak stability condition, an equivalence can be established between rest points of the evolutionary dynamics and MSNE for imitative dynamics.

\begin{corollary}
	Under Assumptions~\ref{ass:cont}--\ref{ass:noise}, consider an imitative or pairwise comparison revision protocol $\rho^c$ for each class $c\in[C]$. Let $\mu^\star$ be a rest point of the evolutionary dynamics \eqref{eq:ODE_mu_ev}. If a solution trajectory of \eqref{eq:ODE_mu_ev} with an interior initial condition $\mu(0) \in \mathrm{int}(X)$ converges to $\mu^\star$, then $\mu^\star$ is a MSNE.
\end{corollary}

\subsection{Local evolutionary stability of strict MSNE}

We study the local stability of a MSNE $\mu^\star$ that has the property of having mass on a single policy that achieves a strictly higher payoff than all other policies. Such a MSNE is called a \emph{strict MSNE}, which is formally defined as follows.

\begin{definition}[Strict MSNE]\label{def:strict_MSNE}
	Under Assumption~\ref{ass:noise}, a MSNE $\mu$ is said to be a strict MSNE if for all $c\in [C]$ and all $u \in \UDc$
	\begin{equation*}
		\sum_{s\in \Scal^c}\!\mu^c[s,u] >0\! \implies \! \left( F^{c,s}_u(\mu) \!>\!  F^{c,s}_v(\mu) \; \forall v \neq u \, \forall s\in \Scal^c \right)\!.
	\end{equation*}
\end{definition}

The evolutionary stability analysis for a strict MSNE is simpler when compared to a generic MSNE and allows to establish stability results under weaker conditions. In the following result, local asymptotic stability of a strict MSNE is established for the whole class of imitative, separable excess payoff, and pairwise comparison revision protocols. 

\begin{theorem}\label{th:stability_strict}
	Under Assumptions~\ref{ass:cont}--\ref{ass:noise}, consider an imitative, separable excess payoff, or pairwise comparison revision protocol $\rho^c$ for each class $c\in [C]$ and let $\mu^\star$ be a strict MSNE. Then, $\mu^\star$ is locally asymptotically stable under the evolutionary dynamics \eqref{eq:ODE_mu_ev}.
\end{theorem}
\begin{proof}
	\ifextendedversion
	See Appendix~\ref{sec:proof_stability_strict}.
	\fi
\end{proof}

%% file: section/MAC.tex
\section{Illustrative Example}\label{sec:MAC}

We illustrate the results of this paper resorting to an application of a medium access game (MAC) between mobile terminals competing for a common wireless channel \cite{WiecekAltmanEtAl2011}.

\subsection{Model}

The model is formally characterized by: 
\begin{itemize}
	\item \emph{Time}: Each player makes a decision each time a Poisson clock with rate $\Rd$ rings. 
	\item \emph{States}: There are four states $\Scal=\{\sE, \sAE,\sAF,\sF\}$ corresponding to four battery levels.
	\item\emph{Actions}: There are three actions $\Acal = \{\aN,\aL,\aH\}$ corresponding to not transmitting, transmitting at low power, and transmitting at high power. We choose $\Acal(\sE) = \{0\}$, $\Acal(\sAE) = \{\aL\}$, $\Acal(\sAF) = \{\aL,\aH\}$, and $\Acal(\sF) = \{\aL,\aH\}$. The transmission powers of actions $\aN$, $\aL$, and $\aH$ are denoted respectively by $P_\aN = 0$, $P_\aL$ and $P_\aH$, which satisfy $0<P_\aL<P_\aH$.
	\item\emph{State transitions}: When a player takes action $\aN$ in state $\sE$ the battery level will be recharged and transition to state $\sF$ with probability $p_\sF$ and will remain at $\sE$ with probability $1-p_\sF$. When a player chooses $a \in \{\aL,\aH\}$, the probability of transitioning to the next lower battery state is $\alpha P_a +\gamma$ and of staying in the same energy level is $1-\alpha P_a -\gamma$. Here, $\alpha>0$ and $\gamma>0$ are constants that model the energy consumption due to the transmission of the message and due to other activities, respectively. These constants must satisfy $\alpha P_\aH + \gamma \leq 1$.
	\item \emph{Single-stage reward}: The single-stage reward of a player in state $s$ playing action $a$ when the state-action distribution of the population is $\mu_{\Scal\times \Acal}\in X_{\Scal\times \Acal}$ is the expected signal to interference and noise ratio given by
	\begin{equation*}
		r(s,a,\mu_{\Scal \times \Acal}) \!=\! \frac{P_a}{\sigma^2 + \Rd T\bar C\!\!\!\!\!\sum\limits_{a^\prime \in \{\aL,\aH\}}\!\!\!\!\! P_{a^\prime}\mu_{\Scal \times \Acal}[\Scal,a^\prime]} -\bar{\beta} P_a,
	\end{equation*}
	where $\sigma, \bar C$, and $\bar{\beta}$ are constants whose physical interpretation is described in \cite{WiecekAltmanEtAl2011}, and $T$ is the duration of the transmission of a message.
\end{itemize}

There are four deterministic policies, which we denote by $\UD = \{u_1,u_2,u_3,u_4\}$. These policies are characterized by 
\begin{equation}\label{eq:mac_policies}
	\small\begin{split}
		u_1(\sE) &= u_2(\sE) = u_3(\sE) = u_4(\sE) = \delta_{\aN}(a)\\
		u_1(\sAE) &= u_2(\sAE) = u_3(\sAE) = u_4(\sAE) = \delta_{\aL}(a)\\
		u_1(\sAF) &= u_2(\sAF) = \delta_{\aL}(a), \quad  u_3(\sAF) = u_4(\sAF) = \delta_{\aH}(a)\\
		u_1(\sF) &= u_3(\sF) = \delta_{\aL}(a), \quad  u_2(\sF) = u_4(\sF) = \delta_{\aH}(a).
	\end{split}	
\end{equation}

\begin{figure}[t!]
	\centering
	\includegraphics[width=\linewidth]{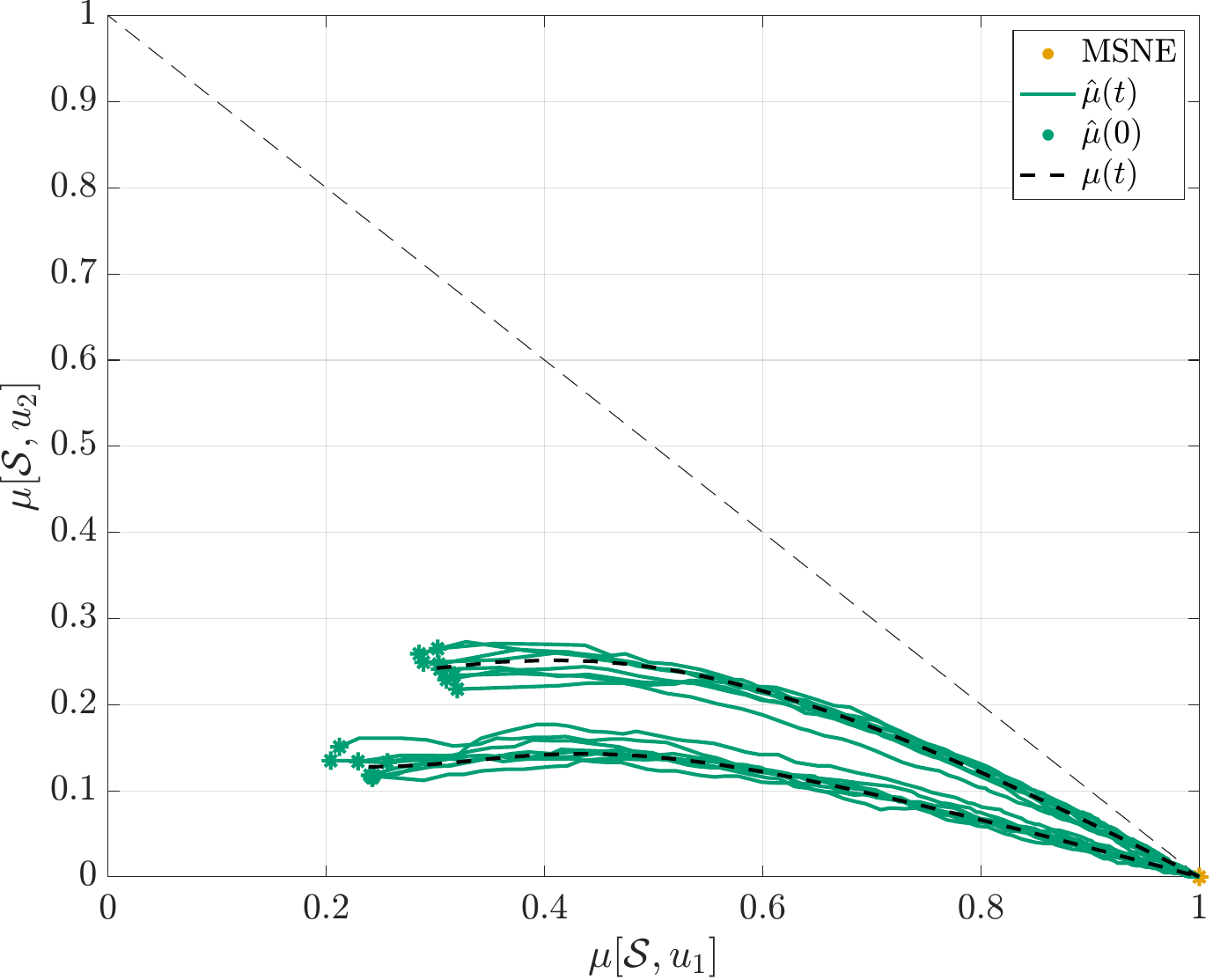}
	\caption{Trajectories of evolutionary dynamics in the illustrative MAC game.}
	\vspace{-0.1cm}
	\label{fig:eg_mac_12}
\end{figure}

We simulate two mean field trajectories of the evolutionary dynamics for an imitative revision protocol with different initial conditions. Moreover, we also simulate several finite-population trajectories for each mean field trajectory, where the player's initial states and policies are drawn randomly from the initial condition of the corresponding mean field trajectory. The finite-population trajectories are simulated with $10^3$ players. Fig.~\ref{fig:eg_mac_12} depicts trajectories along the marginal policy distribution components of policies $u_1$ and $u_2$. First, the trajectories approach a strict MSNE with all the mass in $u_1$, which is a rest-point by Theorem~\ref{th:MSNE_is_rest} and locally asymptotically stable by Theorem~\ref{th:stability_strict}. Second, we can also conclude that the mean field model is a good approximation for the finite-population model as expected from Theorem~\ref{th:approx_ODE_mu_ev}. Code to generate this example for a generic number of battery and transmission power levels is available in an open-access repository at \weblink{https://github.com/fish-tue/evolutionary-mfg-discounted}.

%% file: section/conclusion.tex

\section{Conclusion}

In this paper, we introduced an evolutionary framework for continuous-time, finite-state stochastic dynamic games with many players in the discounted-payoff setting. We showed that the finite-population system can be accurately approximated by a corresponding mean field model. We proposed the mixed stationary Nash equilibrium (MSNE) as a solution concept that has an evolutionary interpretation. We concluded that there is an equivalence between stable MSNE and the rest points of the mean field evolutionary dynamics under two broad families of revision protocols. These results highlight qualitative differences between the discounted-payoff and average-payoff settings. We also concluded that strict MSNE are locally asymptotically stable.

There remain several promising directions for further work in the discounted-payoff setting. First, one could define a finite-population solution concept analogous to the MSNE and establish approximation guarantees relative to the MSNE. This is important because the approximation properties of the mean field evolutionary dynamics in Theorem~\ref{th:approx_ODE_mu_ev} hold only over finite horizons. Second, the evolutionary stability analysis could be extended beyond strict MSNE. In particular, in regimes where revision dynamics evolve significantly more slowly than state dynamics, one may derive conditions on both the structure of the MSNE and the game’s payoff map that ensure local stability of generic MSNE.

%% file: section/proofs_mfg_model.tex
\subsection{Proof of Theorem~\ref{th:approx_ODE_mu_ev}}\label{sec:proof_th_approx_ODE_mu_ev}

First, notice that \eqref{eq:ODE_mu_ev} can be written for all classes $c\in [C]$, all states $s\in \Scal^c$, and all policies $u\in \UDc$ in vector form as an ODE with a vector field $V: X \to T  X$, where $ T  X$ denotes the tangent bundle of $X$. Second, notice that for all $c\in [C]$, all $s\in \Scal^c$, and all $u\in \UDc$, $J^c(u,s,\mu_{\Scal \times \Acal})$, as defined in \eqref{eq:def_J}, can be written as a linear combination of a finite number of single-stage reward functions. Therefore, due to Assumption~\ref{ass:cont}, $J^c(u,s,\mu_{\Scal \times \Acal})$ is Lipschitz continuous w.r.t.\ $\mu$. On the one hand, $f^{c,\mathrm{d}}_{s,u}(\mu)$ defined in \eqref{eq:fd_fr} is linear for all $c\in [C]$, all $s\in \Scal^c$, and all $u\in \UDc$ and, thus, globally Lipschitz continuous. On the other hand, albeit not as obvious, substituting the expressions for imitative, excess payoff, and pairwise comparison revision protocols in $f^{c,\mathrm{r}}_{s,u}(\mu)$ defined in \eqref{eq:fd_fr}, one can also conclude that it is globally Lipschitz continuous for all $c\in [C]$, all $s\in \Scal^c$, and all $u\in \UDc$. Hence, $V(\mu)$ is Lipschitz continuous w.r.t.\ $\mu$. Under these conditions, since $X$ is convex and compact, existence and uniqueness follows from an extension of the Picard-Lindel\"{o}f Theorem to compact convex spaces \cite[Theorem~5.7]{Smirnov2002}\cite[Theorem~4.A.5]{Sandholm2010} and Lipschitz continuity follows from Gr\"onwall's Inequality \cite[Theorem~4.A.3]{Sandholm2010}. Since $V(\mu)$ is Lipschitz continuous and by Assumption~\ref{ass:rev_protocol}, the approximation guarantee w.r.t.\ the finite-population game follows immediately from the application of Kurtz's Theorem \cite[Theorems~10.2.1 and~10.2.3]{Sandholm2010}.

%% file: section/proofs_MSNE.tex

\ifextendedversion
\subsection{Proof of Theorem~\ref{th:existence_MSNE}}\label{sec:proof_th_existence_MSNE}
\fi

The proof relies on Berge's Maximum Theorem and Kakutani's fixed point theorem, whose versions employed in the proof are presented as follows. Set-valued maps are denoted by capital Greek letters and double arrows $\rightrightarrows$.

\begin{theorem}[{\cite[Theorem~17.31]{AliprantisBorder2006}}]\label{th:berge}
Let $\Theta \subset \R^m$, $X\subset \R^n$, and let $C : \Theta \rightrightarrows X$ be a non-empty compact-valued correspondence. Let $f: X \times \Theta \to \R$ be a continuous function. Define the correspondence $\Gamma: \Theta \rightrightarrows X$ by $\Gamma(\theta) = \argmax_{x \in C(\theta)} f(x,\theta)$. If $C$ is continuous (upper and lower hemicontinuous) at $x$, then $\Gamma$ is nonempty, compact, and upper hemicontinuous at $x$.
\end{theorem}

\begin{theorem}[{\cite[Theorem~3.9 ]{BeavisDobbs1990}, \cite[Lemma 20.1]{OsborneRubinstein1994}}]\label{th:kakutani} Let $X \in \R^n$ be compact and convex and let $\Gamma: X \rightrightarrows X$ be a upper hemicontinuous set-valued correspondence such that for all $x\in X$ the set $\Gamma(x)$ is nonempty, compact, and convex. Then, there exists $x^\star \in X$ such that $x^\star \in \Gamma(x^\star)$.
\end{theorem}

We construct a set-valued map and show that a fixed point exists and is a MSNE. Define the space of marginal policy distributions of a class $c\in [C]$ as $X^c_{\UD}:= \{\nu \in \Rnn^{n^c}: \ones^\top\nu = m^c\}$ and define $X_{\UD}:= \bigtimes_{c\in[C]} X^c_{\UD}$. We say that $x\in X_{\UD}$ is a marginal policy distribution and we denote the corresponding policy distribution of a class $c\in [C]$ by $x^c \in X^c_{\UD}$. Define the space of marginal state distributions for each policy of a class  $c\in [C]$ as $X^c_{\Scal}:= \bigtimes_{u\in\UDc} \Pcal(\Scal^c)$ and define $X_{\Scal}:= \bigtimes_{c\in[C]} X^c_{\Scal}$. We say that $\eta \in X_{\Scal}$ is a collection of state distributions for each policy and we denote the corresponding state distribution of class $c\in [C]$ and of policy $u\in \UDc$ by $\eta^{c,u}\in \Pcal(\Scal^c)$. 

First, define the set-valued correspondence $\Gamma_1 : X_{\UD} \times  X_{\Scal} \rightrightarrows  X_{\Scal}$ as 
\begin{equation*}
	\begin{split}
		&\Gamma_1(x,\eta) := \Bigg\{ \nu \in X_{\Scal}  \; \big| \nu^{c,u}(s) = \\
		&\!\sum_{s^\prime \!\in \Scal^c} \!\sum_{a^\prime \!\in \Acal^c}\!\!\!\!\phi^c(s|s^\prime\!,a^\prime)u(a^\prime|s^\prime)\eta^{c,u}(s^\prime),\forall c\!\in\! \![C] \forall s \!\in\! \Scal^c \forall u\!\in \!\UDc \!\!\Bigg\}\!.
	\end{split}
\end{equation*}
Notice that for any $(x,\eta)\in X_{\UD} \times  X_{\Scal}$,  $\Gamma_1$ is single-valued (thus, non-empty, compact, and convex) and upper hemicontinuous at any $(x,\eta)\in X_{\UD} \times  X_{\Scal}$. 

Second, define $\Theta  =  X_{\UD} \times  X_{\Scal}$ and for each $c\in [C]$ the set-valued constant correspondence $C^c:\Theta \rightrightarrows  X^c_{\UD}$ as $C^c(x,\eta):=X^c_{\UD}$, which is non-empty, compact, and continuous. For each $c\in [C]$, define $f^c:X^c_{\UD} \times \Theta \to \R$ as $f^c(x^c,y,\nu):= \sum_{u\in \UDc}x^c[u]J^c(u,\eta^c_\mathrm{unif} ,\mu_{\Scal \times \Acal}(y,\nu))$, 
where $\eta^c_\mathrm{unif} \in \Pcal(\Scal^c)$ is the uniform distribution on $\Scal^c$, and $\mu_{\Scal \times \Acal}(x,\eta) \in X_{\Scal\times \Acal}$ is characterized by $\mu^c_{\Scal \times \Acal}(x,\eta)[s,a] = \sum_{u\in \UDc}x^c[u]\eta^{c,u}(s)u(a|s)$ for all $c\in[C]$, all $s\in \Scal^c$, and all $a\in \Acal^c$. 
Notice that for all $u\in \UDc$, $J^c(u,\eta^c_\mathrm{unif} ,\mu_{\Scal \times \Acal}(y,\nu))$, as defined in \eqref{eq:def_J}, can be written as a linear combination of a finite number of single-stage reward functions. Therefore, due to Assumption~\ref{ass:cont}, $J^c(u,\eta^c_\mathrm{unif} ,\mu_{\Scal \times \Acal}(y,\nu))$ is continuous w.r.t.\ $(y,\nu)$. As a result, $f^c$ is continuous on $ X^c_{\UD}\times \Theta$. For each $c\in[C]$, define the set-valued correspondence $\Gamma^c_2 : \Theta \rightrightarrows X^c_{\UD}$ as  $\Gamma_2^c(\theta) = \argmax_{x^c \in C^c(\theta)} f^c(x^c,\theta)$. Since we are in the conditions of Theorem~\ref{th:berge}, it follows that $\Gamma_2^c$ is nonempty, compact, and upper hemicontinuous. Furthermore, one can also show that $\Gamma^c_2(\nu_{\UD})$ is convex for all $\theta \in \Theta$ since $C^c(\theta)$ is convex and $x^c$ defines a convex combination of elements parametrized by $\theta$ in $f^c(x^c,\theta)$. 

Third, define $\Gamma$ as $\gamma := \bigtimes_{c\in [C]}\Gamma^c_2 \times \Gamma_1$. Thus, $\Gamma$ is nonempty, compact, convex, and upper hemicontinuous. It follows immediately from Theorem~\ref{th:kakutani}, that $\Gamma$ admits a fixed point. The following proposition shows that a fixed-point of $\Gamma$ is a MSNE, which concludes the proof.

\begin{proposition}\label{prop:proof_existence_fixed_point}
	If $(x,\eta) \in  X_{\UD} \times  X_{\Scal}$ is a fixed point of $\Gamma$, i.e., $(x,\eta) \in \Gamma(x,\eta)$, then $\mu \in X$ characterized by $\mu^c[s,u] := x^c[u]\eta^{c,u}(s)$ for all $c\in[C]$, all $s\in \Scal^c$, and all $u\in \UDc$ is a MSNE.
\end{proposition}

\begin{proof}
	Let $(x,\eta) \in  X_{\UD} \times  X_{\Scal}$ be a fixed point of $\Gamma$. The goal is to show that $\mu \in X$ defined in the statement of the proposition is a MSNE according to Definition~\ref{def:MSNE}. First, we check \eqref{eq:MSNE_s}. Notice that for all $c\in [C]$, all $s\in \Scal^c$, all $u\in \UDc$
	\begin{equation*}
		\begin{split}
			\mu^c[s,u]  & =  x^c[u] \eta^{c,u}(s)\\
			&  =   x^c[u] \sum_{s^\prime \in \Scal^c} \sum_{a^\prime \in \Acal^c}\phi^c(s|s^\prime,a^\prime)u(a^\prime|s^\prime)\eta^{c,u}(s^\prime) \\
			& = \sum_{s^\prime \in \Scal^c} \sum_{a^\prime \in \Acal^c}\phi^c(s|s^\prime,a^\prime)u(a^\prime|s^\prime)\mu^c[s^\prime,u].
		\end{split}
	\end{equation*}
	Therefore, $\mu$ satisfies \eqref{eq:MSNE_s}. Second, we check \eqref{eq:MSNE_u}. By the definition of a fixed point of $\Gamma$, it follows that for all $c\in [C]$ and all $u,v\in \UDc$
	\begin{equation}\label{eq:proof_opt_u}
		\!x^c[u] \!>\!0\! \implies\! J^c(u,\eta^c_\mathrm{unif} ,\mu_{\Scal \times \Acal}) \!\geq\! J^c(v,\eta^c_\mathrm{unif} ,\mu_{\Scal \times \Acal}).\!
	\end{equation}
	Since $\mu^c[s,u] := x^c[u]\eta^{c,u}(s)$ it follows that $\mu^c[s,u] >0 \implies x^c[u] >0$. Moreover, the cost function $J$, defined in \eqref{eq:def_J}, defines a finite Markov decision process (MDP)  with a discounted infinite-horizon cost, and \eqref{eq:proof_opt_u} states that any policy in the support of $x^c$ is optimal starting for a state distribution $\eta^c_\mathrm{unif}$ that places mass on every state. As a result, from standard optimality results of finite MDPs \cite[Theorem~6.2.7]{Puterman1994}, any policy in the support of $x^c$  is optimal starting from any state $s\in \Scal^c$. Combining these two observations yields
	\begin{equation*}
		\mu^c[s,u] \!>\!0 \!\implies\!\!J^c(u,s ,\mu_{\Scal \times \Acal}(x,\!\eta)\!) \!\geq \! J^c(v,s ,\mu_{\Scal \times \Acal}(x,\!\eta)\!)
	\end{equation*} 
	 for all $c\in [C]$, all $s\in \Scal^c$, all $u,v\in \UDc$. Hence, $\mu$ also satisfies \eqref{eq:MSNE_u}, which concludes the proof.
\end{proof}

%% file: section/proofs_MSNE_rest.tex

\subsection{Proof of Theorem~\ref{th:MSNE_is_rest}}\label{sec:proof_th_MSNE_is_rest}

This proof relies mainly on the following proposition.

\begin{proposition}\label{prop:PC}
	Let $\rho^c$ be a imitative, excess payoff, or pairwise comparison revision protocol. Define $V^c_u(\sigma,F) =  \sum_{u^\prime \in \UDc}  \sigma[u^\prime] \rho^c_{u^\prime u}(F,\sigma) - \sigma[u] \sum_{u^\prime \in \UDc}  \rho^c_{uu^\prime}(F,\sigma)$, where $ F\in \R^{n^c}$ and $\sigma \in \Delta_{\UD}^c$. Then, the revision protocol $\rho^c$ satisfies positive correlation, i.e., $V^c(\sigma,F):=\col(V^c_u(\sigma,F),u\in \UDc) \neq \zeros \implies \sum_{u\in \UDc} V^c_u(\sigma,F)F^c_u >0$, for all $F\in \R^{n^c}$, and all $\sigma \in \Delta^c_{\UD}$.
\end{proposition}
\begin{proof}
	Well-known results in evolutionary game theory \cite[Theorems~5.4.9, 5.5.2, and~5.6.2]{Sandholm2010} establish that imitative, excess payoff, and pairwise comparison revision protocols satisfy positive correlation.
\end{proof}

Consider any MSNE $\mu \in X$. First, from \eqref{eq:fd_fr} and \eqref{eq:MSNE_s}, it follows that $f^{c,\mathrm{d}}_{s,u}(\mu) = 0$ for all $c \in [C]$, all $s\in \Scal^c$, and all $u\in \UDc$. Second, notice that $f^{c,\mathrm{r}}_{s,u}(\mu) = V_u^c(\mu^c[s,\cdot],F^{c,s}(\mu))$ for all $c \in [C]$, all $s\in \Scal^c$, and all $u\in \UDc$, where $V^c_u$ is defined in Proposition~\ref{prop:PC}. From \eqref{eq:MSNE_u} in the definition of MSNE it follows that $(\sigma -\mu^c[s,\cdot])^\top F^{c,s}(\mu) \leq 0$ for all $c\in [C]$, all $s\in \Scal^c$, and all $\sigma \in \{x\in \Delta^c_{\UD}: \ones^\top x  = \ones^\top \mu^c[s,\cdot]\}$. Choosing, e.g., $\sigma =  \mu^c[s,\cdot] +  V^c(\mu^c[s,\cdot])$ it follows that $V^c(\mu^c[s,\cdot],F^{c,s}(\mu))^\top F^{c,s}(\mu) \leq 0$. Proposition~\ref{prop:PC} implies immediately that $V^c(\mu^c[s,\cdot],F^{c,s}(\mu))  = \zeros$ for all $c\in [C]$ and all $s\in \Scal^c$. Hence, $f^{c,\mathrm{r}}_{s,u}(\mu) = 0$ for all $c \in [C]$, all $s\in \Scal^c$, and all $u\in \UDc$, which concludes the proof.

\subsection{Proof of Theorem~\ref{th:rest_is_MSNE}}\label{sec:proof_rest_is_MSNE}

This proof relies on showing the result for any revision protocols that satisfy a sign preservation property, which is defined in what follows. Notice that pairwise comparison revision protocols satisfy sign preservation by definition.

\begin{definition}[Sign Preservation]
	A revision protocol $\rho^c$ is said to satisfy sign preservation if $\sign(\rho^c_{uv}(F,\sigma)) = \sign(\max(0,F_v-F_u)),  \forall F \in \R^{n^c}\;\forall \sigma \in \Delta_{\UD}^c \;\forall u,v \in \UDc$.
\end{definition}

\begin{proposition}\label{prop:not_rest_not_MSE}
	Consider any revision protocol $\rho^c$ for each class $c\in [C]$ that satisfies sign preservation. If $\mu \in X$ is a rest point of the evolutionary dynamics \eqref{eq:ODE_mu_ev} and $\mu$ is a not a MSNE, then there exists $c \in [C]$, $s^\dagger \in \Scal^{c}$, and $u^\dagger \in \UDc$ such that $\mu^{c}[s^\dagger,u^\dagger]>0$ and
	\begin{equation}\label{eq:Vsu_neq_0}
		\begin{split}
			&\Rdc \!\!\sum_{s^\prime\! \in \Scal^c} \! \sum_{a^\prime \!\in \Acal^c}\!\! \phi^c(s|s^\prime,a^\prime)u^\dagger(a^\prime|s^\prime)\mu^c[s^\prime,u^\dagger] - \Rdc\mu^c[s^\dagger,u^\dagger] \!\!\!\!\!\!\!\!\!\!\!\!\!\!\\
			\!\!=&- \!\!\sum_{u^\prime \in \UDc}  \mu^c[s^\dagger,u^\prime] \rho^c_{u^\prime u^\dagger}(F^{c,s^\dagger}(\mu),\mu^c[s^\dagger,\cdot])  \\
			&\quad \quad \; +\mu^c[s^\dagger,u^\dagger] \sum_{u^\prime \in \UDc}  \rho^c_{u^\dagger u^\prime}(F^{c,s^\dagger}(\mu),\mu^c[s^\dagger,\cdot]) \neq 0.\!\!\!\!
		\end{split}
	\end{equation}
\end{proposition}
\begin{proof}
	By the definition of rest point of the evolutionary dynamics in \eqref{eq:ODE_mu_ev}, $\dot\mu^c[s,u] = 0$ for all $c\in [C]$, all $s\in \Scal^c$, and all $u\in \UDc$. Then, either (i)
	\begin{equation}\label{eq:Vsu_eq_0_1}
			\!\Rdc \!\! \sum_{s^\prime \! \in \Scal^c} \! \sum_{a^\prime \! \in \Acal^c} \!\!\!\phi^c(s|s^\prime\!,a^\prime)u(a^\prime|s^\prime)\mu^c[s^\prime,u] \!-\! \Rdc\mu^c[s,u] \!=\! 0
	\end{equation}
	and
	\begin{equation}\label{eq:Vsu_eq_0_2}
		\begin{split}
			\sum_{u^\prime \in \UDc} \mu^c&[s,u^\prime] \rho^c_{u^\prime u}(F^{c,s}(\mu),\mu^c[s,\cdot]) \\
			&- \mu^c[s,u] \sum_{u^\prime \in \UDc}  \rho^c_{u u^\prime}(F^{c,s}(\mu),\mu^c[s,\cdot]) = 0
		\end{split}
	\end{equation}
	for all $c\in [C]$, all $s \in \Scal$, and all $u\in \UDc$; or (ii)~there exists $c\in[C]$, $s^\dagger \in \Scal^c$ and $u^\dagger\in \UDc$ such that \eqref{eq:Vsu_neq_0} is satisfied.
	Since statements~(i) and~(ii) are complementary, we show statement~(ii) by proving that statement~(i) implies a contradiction. Assume, by contradiction, that statement~(i) is true. Condition \eqref{eq:Vsu_eq_0_2} establishes stationarity of the evolutionary dynamics with vanishing state dynamics. It is known that under sign preservation without state dynamics \cite[Lemma~5.6.4]{Sandholm2010} that \eqref{eq:Vsu_eq_0_2} implies that $\mu$ satisfies condition \eqref{eq:MSNE_u} in Definition~\ref{def:MSNE} of a MSNE. In addition, from \eqref{eq:Vsu_eq_0_1}, it follows that $\mu$ satisfies condition \eqref{eq:MSNE_s} in Definition~\ref{def:MSNE} of a MSNE. One concludes that $\mu$ is a MSNE, which is a contradiction, thereby proving statement~(ii). 
	We also prove by contradiction that given $c\in [C]$, $s^\dagger \in \Scal^c$ and $u^\dagger\in \UDc$ such that \eqref{eq:Vsu_neq_0} is satisfied, it follows that $\mu^c[s^\dagger,u^\dagger]>0$. Assume by contradiction that $\mu^c[s^\dagger,u^\dagger]=0$. It follows from \eqref{eq:Vsu_neq_0} that 
	$\Rdc \sum_{s^\prime \in \Scal^c} \sum_{a^\prime \in \Acal^c}\phi^c(s|s^\prime,a^\prime)u(a^\prime|s^\prime)\mu^c[s^\prime,u] = - \sum_{u^\prime \in \UDc}  \mu^c[s^\dagger,u^\prime] \rho^c_{u^\prime u^\dagger}(F^{c,s^\dagger}(\mu),\mu^c[s^\dagger,\cdot])$, whose left hand side is nonnegative and right hand side is nonpositive. It follows that both sides must be null, which is a contradiction. 
\end{proof}

Finally, we prove Theorem~\ref{th:rest_is_MSNE} by contradiction. Assume that $\mu \in X$ is a rest point of the evolutionary dynamics \eqref{eq:ODE_mu_ev} and that $\mu$ is a not a MSNE. From Proposition~\ref{prop:not_rest_not_MSE}, it follows that there exist $c\in [C]$, $s^\dagger \in \Scal^c$ and $u^\dagger\in \UDc$ such that $\mu^c[s^\dagger,u^\dagger]>0$ and \eqref{eq:Vsu_neq_0} is satisfied. First, we consider the case where $u^\dagger$ is not optimal (at least from) state $s^\dagger$. In that case, there exists $u^\star \in \UDc \setminus \{u^\dagger\}$ such that $F^{c,s}_{u^\star}(\mu) \geq F^{c,s}_{v}(\mu)$ for all $v\in \UDc$ and for all $s\in\Scal^c$ and $F^{c,s^\dagger}_{u^\star}(\mu) > F^{c,s^\dagger}_{u^\dagger}(\mu)$. Since $\rho^c$ satisfies sign preservation and $\mu^c[s^\dagger,u^\dagger]>0$, $\sum_{s\in \Scal^c} \mu^c[s,u^\dagger]\rho^c_{u^\dagger,u^\star}(F^{c,s}(\mu),\mu^c[s,\cdot]) >0$, thus $\sum_{s\in \Scal^c}\sum_{u^\prime\in \UDc} \mu^c[s,u^\prime]\rho^c_{u^\prime,u^\star}(F^{c,s}(\mu),\mu^c[s,\cdot]) >0$.
Moreover, $\sum_{s\in \Scal^c} \mu^c[s,u^\star] \sum_{u^\prime \in \UDc}  \rho^c_{u^\star u^\prime}(F^{c,s}(\mu),\mu^c[s,\cdot]) = 0$. Hence, $\sum_{s\in \Scal^c}\dot{\mu}^c[s,u^\star] >0$, which is a contradiction with the hypothesis that $\mu$ is a stationary point of \eqref{eq:ODE_mu_ev}.
Second, we consider the case where $u^\dagger$ is optimal for state $s^\dagger$, i.e., $F^{c,s^\dagger}_{u^\dagger}(\mu) \geq F^{c,s^\dagger}_v(\mu)$ for all $v\in \UDc$. 
Since $\rho^c$ satisfies sign preservation,  $\rho^c_{u^\dagger u^\prime}(F^{c,s^\dagger}(\mu),\mu^c[s^\dagger,\cdot]) = 0$ for all $u^\prime \in \UDc$. As a result, from \eqref{eq:Vsu_neq_0}, $- \sum_{u^\prime \in \UDc}  \mu^c[s^\dagger,u^\prime] \rho^c_{u^\prime u^\dagger}(F^{c,s^\dagger}(\mu),\mu^c[s^\dagger,\cdot])  \neq 0$. Therefore, there is $u^\circ \in \UDc$ such that $\mu^c[s^\dagger,u^\circ]>0$ and, since $\rho^c$ satisfies sign preservation, $F^{c,s^\dagger}_{u^\circ}(\mu) < F^{c,s^\dagger}_{u^\dagger}(\mu)$. Employing the same analysis of the first case, but now for policy $u^\circ$, leads to a contradiction, which concludes the proof.

%% file: section/evolutionary_stability_proofs.tex
\section{Proofs of Section~\ref{sec:local_ev_stability}}

\subsection{Proof of Theorem~\ref{th:imitative_rest_unstable}}\label{sec:proof_imitative_rest_unstable}

\begin{proposition}\label{prop:imitative_monotone_growth_rates}
	Consider an imitative revision protocol $\rho^c$ defined as in Definition~\ref{def:imitative}. Then, the evolutionary dynamics \eqref{eq:ODE_mu_ev} can be written as
	\begin{equation}\label{eq:imitative_ev_dyn}
		\begin{split}
			\dot{\mu}^c[s,u] = \mu^c[s,u]G_u^{c,s}(\mu) +  f^{c,\mathrm{d}}_{s,u}(\mu)
		\end{split}
	\end{equation}
	for all $c\in[C]$, all $s\in \Scal^c$, and all $u\in \UDc$ where $G_u^{c,s}(\mu)$ is called a growth rate and is defined as
	\begin{equation}\label{eq:imitative_growth_rate}
		\begin{split}
					G_u^{c,s}(\mu) = \sum_{u^\prime \in \UDc}& \frac{\mu^c[s,u^\prime]}{\ones^\top\mu^c[s,\cdot]}\Big(r^c_{u^\prime u}(F^{c,s}(\mu),\mu^c[s,\cdot])\\&-r^c_{u u^\prime}(F^{c,s}(\mu),\mu^c[s,\cdot])\Big) 
		\end{split}
	\end{equation}
	if $\ones^\top\mu^c[s,\cdot]>0$ and is null otherwise.
 Furthermore, the growth rates are monotonous, i.e., for all $c\in [C]$, all $s\in \Scal$, and all $u,v\in \UDc$ if $\ones^\top\mu^c[s,\cdot]>0$ then $F_u^{c,s}(\mu) \geq F_v^{c,s}(\mu) \iff G_u^{c,s}(\mu) \geq G_v^{c,s}(\mu)$, and for all $c\in [C]$ and all $s\in \Scal^c\;$ $\sum_{u\in\UDc}\mu^c[s,u]G_u^{c,s}(\mu) = 0$. Moreover, if $\mu^\star$ is a MSNE, then $G_u^{c,s}(\mu^\star) = 0$  for all $c\in [C]$, all $s\in \Scal^c$, and all $u\in \argmax_{u^\prime \in \UDc} F^{c,s}_{u^\prime}(\mu^\star)$ and $G_u^{c,s}(\mu^\star) < 0$ for all $c\in [C]$, all $s\in \Scal$, and all $u\notin \argmax_{u^\prime \in \UDc} F_{u^\prime}^{c,s}(\mu^\star)$.
\end{proposition}
\begin{proof}
	Using Definition~\ref{def:imitative} in \eqref{eq:ODE_mu_ev} yields \eqref{eq:imitative_ev_dyn} after some algebraic manipulation.  The monotonicity of the growth rates follows immediately from Definition~\ref{def:imitative} and \eqref{eq:imitative_growth_rate}. The equality $\sum_{u\in\UDc}\mu^c[s,u]G_u^{c,s}(\mu) = 0$ follows from the observation that for class $c\in [C]$ and each state $s\in \Scal^c$, the revision flows of all policies, which are depicted in \eqref{eq:ODE_mu_ev}, are null, i.e., $ \sum_{u\in\UDc} f^{c,\mathrm{r}}_{s,u}(\mu) = 0$. The last statement follows from the fact that if $\mu^\star$ is a MSNE then (i)~for all $c\in [C]$, all $s\in \Scal^c$, and all $u \in \UDc$ $f^{c,\mathrm{d}}_{s,u}(\mu^\star) = 0$ by Definition~\ref{def:MSNE}; (ii)~for all $c\in [C]$ and all $s\in \Scal^c$ at least one policy $u\in \argmax_{u^\prime \in \UDc} F^{s,c}_{u^\prime}(\mu^\star)$ is in the support of $\mu^{c\star}$, by Definition~\ref{def:MSNE} and Assumption~\ref{ass:noise}; and (iii)~$\mu^\star$ is a rest point of  \eqref{eq:imitative_ev_dyn} by Theorem~\ref{th:MSNE_is_rest}. As a result, for all $c\in [C]$ and all $s\in \Scal^c$ there is $u\in \argmax_{u^\prime \in \UDc} F_{u^\prime}^{c,s}(\mu^\star)$ such that $\mu^{c \star}[s,u] >0$ and $\mu^{c\star}[s,u]G^{c,s}_u(\mu^\star) = 0$, which implies $G^{c,s}_u(\mu^\star) = 0$. By monotonicity of the growth rates and the fact that $\ones^\top\mu^{c\star}[s,\cdot]>0$ for all $c\in [C]$ and all $s\in \Scal^c$ by Assumption~\ref{ass:noise}, for all $c\in[C]$ and all $s\in \Scal^c$ all policies in $\argmax_{u^\prime \in \UDc} F_{u^\prime}^{c,s}(\mu^\star)$ have null growth rate and all other policies have strictly negative growth rates.
\end{proof}

\begin{proposition}\label{prop:imitative_nonMSNE_rest}
	Consider that at least one class uses an imitative revision protocol defined as in Definition~\ref{def:imitative}. Let $\mu^\star$ be a non-MSNE rest-point of \eqref{eq:ODE_mu_ev}. Then, under Assumption~\ref{ass:noise}, there exists one class $c\in [C]$ that uses an imitative revision protocol for which optimal policies are not in the support of $\mu^{c\star}$ and their growth rates are positive, i.e., there is $c\in[C]$ such that for all $s\in \Scal^c$ $u\in \argmax_{u^\prime \in \UDc}  F^{c,s}_{u^\prime}(\mu^\star) \implies ( \mu^{c\star}[s,u] = 0 \;\land\; G^{c,s}_u(\mu^\star) > 0)$.
\end{proposition}
\begin{proof}
	Since $\mu^\star$ is, by hypothesis, a rest point of \eqref{eq:ODE_mu_ev} and is not a MSNE, then by Definition~\ref{def:MSNE} and Assumption~\ref{ass:noise}, at least one of the following statements is false:
	\begin{enumerate}
		\renewcommand{\labelenumi}{(\roman{enumi})} 
		\item For all $c\in [C]$ and all $u \in \UDc$, $\sum\nolimits_{s\in \Scal^c}\mu^{c\star}[s,u] >0 \implies F^{c,s}_u(\mu^{\star}) \geq F_{v}^{c,s}(\mu^\star)$ for all $s\in \Scal^c$ and $v\in\UDc$;
		\item For all $c\in [C]$, all $s\in \Scal^c$, and all $u\in \UDc$, $f^{c,\mathrm{d}}_{s,u}(\mu^\star) = 0$.
	\end{enumerate}
	First, we proceed to prove that statement~(i) is always false. Assume, by contradiction, that statement~(i) is true, i.e., all the mass is placed on optimal policies. As a result, since  $\mu^\star$ is not a MSNE, statement~(ii) must be false. With a stationary state-policy distribution $\mu^\star$ the policy decision is a MDP, thus, from standard optimality results of finite MDPs \cite[Theorem~6.2.7]{Puterman1994} a deterministic optimal policy exists and it is optimal from every state. Since statement~(i) is true, all the mass of $\mu^\star$ is on optimal policies. As a result, since statement (ii) is false and all mass is on optimal policies, there is $c\in [C]$ and $u^\star \in \UDc$ such that $F_{u^\star }^{c,s}(\mu^\star) =\max_{u^\prime\in \UDc} F_{u^\prime}^{c,s}(\mu^\star)$ and   $\mu^{c\star}[s,u^\star]>0$ for all $s\in \Scal$ and  $f^{c,\mathrm{d}}_{s,u^\star}(\mu^\star) \neq 0$ for some $s\in \Scal^c$. Since $\sum_{s\in \Scal^c} f^{c,\mathrm{d}}_{s,u^\star}(\mu^\star) = 0$
	and $\mu^\star$ is a rest point of \eqref{eq:imitative_ev_dyn}, then there is $s^\star \in \Scal^c$ such that $f^{c,\mathrm{d}}_{s^\star,u^\star}(\mu^\star) >0$ and $G^{c,s^\star}_{u^\star}(\mu^\star) <0$ by \eqref{eq:imitative_ev_dyn} in Proposition~\ref{prop:imitative_monotone_growth_rates}. By the monotonicity of the growth rates from Proposition~\ref{prop:imitative_monotone_growth_rates}, it follows that  $\sum_{u\in\UDc}\mu^c[s^\star,u]G_u^{c,s^\star}(\mu^\star) <0$, which is a contradiction by Proposition~\ref{prop:imitative_monotone_growth_rates}.
	
	Second, we prove the main result of the proposition. Given that statement~(i) is false and by Assumption~\ref{ass:noise}, then there is $c\in [C]$ and a policy $v\in \UDc$ such that for all $s\in \Scal^c$ $\mu^{c\star}[s,v] > 0$ and $F_{v}^{c,s}(\mu^\star) <\max_{u^\prime\in \UDc} F_{u^\prime}^{c,s}(\mu^\star)$. Since $\mu^\star$ is a rest point, one of two cases occurs:
	\begin{enumerate}
		\renewcommand{\labelenumi}{(\alph{enumi})} 
		\item $G^{c,s}_v = 0$ for all $s\in \Scal^c$, i.e., when $f^{c,\mathrm{d}}_{s,v}(\mu^\star) = 0$ for all $s\in \Scal^c$;
		\item There is $l\in \Scal^c$ such that $G^{c,l}_v < 0$. 
	\end{enumerate}
	In case~(b), since $\mu^{c\star}[v,l] >0$ and, by Proposition~\ref{prop:imitative_monotone_growth_rates}, $\sum_{u\in\UDc}\mu^c[s,l]G_u^{c,l}(\mu) = 0$, there exists a policy $v^\prime\in \UDc$ such that $G_{v^\prime}^{c,l}(\mu^\star)>0$. By the monotonicity of the growth rates in Proposition~\ref{prop:imitative_monotone_growth_rates}, it follows that the growth rates are positive for all optimal policies, i.e., for all $s\in \Scal^c$ $u\in \argmax_{u^\prime \in \UDc}  F^{c,s}_{u^\prime}(\mu^\star) \implies G^{c,s}_u(\mu^\star) > 0$. Finally, by Proposition~\ref{prop:imitative_monotone_growth_rates}, since $\sum_{u\in\UDc}\mu^\star[s,u]G_u^{c,s}(\mu^\star) = 0$, optimal policies do not have mass in $\mu^{c\star}$.
\end{proof}

Given a rest point $\mu^\star$, the policy decision is a MDP, so from standard optimality results of finite MDPs \cite[Theorem~6.2.7]{Puterman1994} there is at least one deterministic policy for each class that is optimal from every state of that class. For class $c\in[C]$, define the set of such policies by $\UDcbar =  \argmax_{u^\prime \in \UDc}  F^{c,s}_{u^\prime}(\mu^\star)$ for any state $s\in \Scal^c$. It follows from Proposition~\ref{prop:imitative_nonMSNE_rest} and by continuity of the growth rates defined in Proposition~\ref{prop:imitative_monotone_growth_rates} that in a sufficiently small neighborhood of $\mu^\star$, denoted by $\Ocal_{\mu^\star}$,  for all $c\in [C]$, all $s\in \Scal^c$, all $u\in \UDcbar$ and all $\mu\in \Ocal_{\mu^\star}$, $G^{c,s}_u(\mu) \geq  k > 0$, where $k>0$. Consider a solution trajectory $\{\mu(t)\}_{t\geq 0}$ of \eqref{eq:ODE_mu_ev} with, $\mu(0)\in \Ocal_{\mu^\star}$, and define $\sigma^c(t) = \sum_{u\in \UDcbar}\sum_{s\in \Scal^c} \mu^c[s,u](t)$. Notice that since $\sum_{s\in \Scal^c} f^{c,\mathrm{d}}_{s,u^\star}(\mu^\star) = 0$, 
\begin{equation*}
	\begin{split}
		\dot{\sigma}^c(t) & =  \sum_{u\in \UDcbar}\sum_{s\in \Scal^c} \mu^c[s,u](t)G_u^{c,s}(\mu(t))\\
		& \geq k \sum_{u\in \UDcbar}\sum_{s\in \Scal^c} \mu^c[s,u](t) = k\sigma^c(t).
	\end{split}
\end{equation*}
Since $k>0$, no matter the choice of  $\mu(0)\in \Ocal_{\mu^\star}\setminus \{\mu^\star\}$, any solution trajectory leaves $\Ocal_{\mu^\star}$ in finite time. Therefore, $\mu^\star$ is not a Lyapunov stable equilibrium of \eqref{eq:ODE_mu_ev}. Furthermore, by \eqref{eq:imitative_ev_dyn} and Lipschitz continuity of the vector field of \eqref{eq:ODE_mu_ev} by Theorem~\ref{th:approx_ODE_mu_ev}, it follows that for any initial condition in $X$, the signal of the components of $\sum_{s\in \Scal^c}\mu[s,\cdot](t)$ for all $c\in[C]$ is preserved forward in time. As a result, no solution trajectory of \eqref{eq:ODE_mu_ev} with initial condition in $\mathop{\mathrm{int}}(X)$ converges to $\mu^\star$, which concludes the proof.

\subsection{Proof of Theorem~\ref{th:stability_strict}}\label{sec:proof_stability_strict}

The proof relies on LaSalle's invariance principle~\cite[Chap.~4.2]{Khalil2002}. 
Since the candidate Lyapunov function is not continuously differentiable, we use a generalized derivative called the upper right Dini derivative, which we define in what follows alongside with useful properties.

\begin{definition}[{\cite[Definition A.15]{Bullo2024}}]
	The upper right Dini derivative of a continuous function $f:] a, b[\rightarrow \mathbb{R}$ at a point $t \in] a, b[$ is defined by
	\begin{equation*}
		D^{+} f(t)=\limsup _{\Delta t>0, \Delta t \rightarrow 0} \frac{f(t+\Delta t)-f(t)}{\Delta t}.
	\end{equation*}
\end{definition}

\begin{lemma}[{\cite[Lemma A.16]{Bullo2024}}]\label{lemma:prop_Dini}
	Given a continuous function $f:] a, b[\rightarrow \mathbb{R}$
	\begin{enumerate}[(i)]%
		\item  if $f$ is differentiable at $t \in ] a, b[$, then $D^{+} f(t)=\frac{d}{d t} f(t)$ is the usual derivative of $f$ at $t$;
		\item  if $D^{+} f(t) \leq 0$  for all $t \in ] a, b[$, then $f$ is non-increasing on $] a, b[$.
	\end{enumerate}
\end{lemma}

Before defining a candidate Lyapunov function, two propositions that allow to check the conditions of LaSalle's Theorem \cite[Theorem~A.7]{Bullo2024} are established. Since $\mu^\star$ is a strict MSNE, by Definition~\ref{def:strict_MSNE}, for each class $c\in [C]$ there is a policy $u^{c\star} \in \UDc$ such that $\sum_{s\in \Scal^c}\mu^{c\star}[s,u^\star] = 1$ and $F_{u^{c\star}}^{c,s}(\mu^\star) > F_{v}^{c,s}(\mu^\star)$ for all $v\in \UDc \setminus \{u^{c\star}\}$ and $s\in \Scal^c$. We define a neighborhood $D_\alpha(\mu^\star) \subset X$ of $\mu^\star$ for some $\alpha >0$ as 
\begin{equation*}
	D_\alpha(\mu^\star)  := \left\{\mu\in  X : ||\mu-\mu^\star||_1 \leq \alpha\right\}.
\end{equation*} 

\begin{proposition}\label{prop:strict_propoerties_rev}
	Consider an imitative, separable excess payoff, or pairwise comparison revision protocol for each class $c\in [C]$. Then, there exists $\alpha_1>0$ such that for all $c\in [C]$, all $\mu\in D_{\alpha_1}(\mu^\star)$, and all $s\in \Scal^c$:
	\begin{enumerate}
		\renewcommand{\labelenumi}{(\roman{enumi})} 
		\item $f^{c,\mathrm{r}}_{s,u^{c\star}}(\mu) \geq 0$;
		\item $f^{c,\mathrm{r}}_{s,u^{c\star}}(\mu) = 0 \iff \sum_{v\in \UDc\setminus \{u^{c\star}\}}\mu^c[s,v] = 0$.
	\end{enumerate}
\end{proposition}
\begin{proof}
	First, we consider any class $c\in [C]$ that uses an imitative revision protocol. Since $\mu^\star$ is a MSNE, for all $u \in \UDc$ and all $s\in \Scal^c$ $f^{c,\mathrm{d}}_{s,u}(\mu^\star) = 0$. Therefore, by Theorem~\ref{th:MSNE_is_rest} it follows that $f^{c,\mathrm{d}}_{s,u}(\mu^\star) +f^{c,\mathrm{r}}_{s,u}(\mu^\star) = 0$, and, hence, $f^{c,\mathrm{r}}_{s,u}(\mu^\star) = 0$ for all $u \in \UDc$ and all $s\in \Scal^c$. By Proposition~\ref{prop:imitative_monotone_growth_rates}, it follows that for all $s\in \Scal^c$ $f^{c,\mathrm{r}}_{s,u^{c\star}}(\mu^\star) = G^{c,s}_{u^{c\star}}(\mu^\star)\mu^{c\star}[s,u^{c\star}] = 0$. Therefore, since $\mu^c[s,u^{c\star}] >0$, $G_{u^{c\star}}^{c,s}(\mu^\star) = 0$. By the monotonicity of the growth rates established in Proposition~\ref{prop:imitative_monotone_growth_rates}, it follows that for all $v\in \UDc \setminus \{u^{c\star}\}$ and all $s\in \Scal^c$ $G_{v}^{c,s}(\mu^\star)<0$. Therefore, by continuity, there is $\alpha^c_1>0$ such that $ \mu \in D_{\alpha^c_1}(\mu^\star) \implies G_{v}^{c,s}(\mu)<0 \;\forall v\in \UDc \setminus \{u^{c\star}\} \forall s\in \Scal^c$. By Proposition~\ref{prop:imitative_monotone_growth_rates}, $\sum_{u\in \UDc}G^{c,s}_{u}(\mu)\mu^c[s,u] = 0$, therefore for all $\mu\in D_{\alpha_1^c}(\mu^\star)$  and $s\in \Scal^c$
	\begin{equation}\label{eq:imitative_signa_fr_strict}
		\begin{split}
					f_{s,u^{c\star}}^{c,\mathrm{r}}(\mu) &= G_{u^{c\star}}^{c,s}(\mu)\mu^c[s,u^\star] \\~
					&= -\sum_{v\in \UDc \setminus \{u^{c\star}\}} G_v^{c,s}(\mu)\mu^c[s,v] \geq 0.
		\end{split}
	\end{equation} 
	Furthermore, \eqref{eq:imitative_signa_fr_strict} holds with equality if and only if $\sum_{v\in \UDc\setminus \{u^{c\star}\}}\mu^c[s,v] = 0$.
	
	Second, we consider any class $c$ that uses a separable excess payoff revision protocol. From the expression for the average payoff, it follows that for all $s\in \Scal^c$, all $u\in \UDc \setminus \{u^{c\star}\}$, and all $\mu \in D_\alpha(\mu^\star)$, since $\ones^\top \mu^c[s,\cdot]>0$ for sufficiently small $\alpha>0$ from Assumption~\ref{ass:noise},
	{\allowdisplaybreaks
	\begin{align}\label{eq:aux_avg_payoff} 
		\!\!\!\!\hat{F}^{c,s}_u(\mu) &= F^{c,s}_u(\mu)-\sum_{v\in \UDc}\frac{\mu^c[s,v]}{\ones^\top \mu^c[s,\cdot]}F^{c,s}_v(\mu)\notag \\
		& = F^{c,s}_u(\mu) - \frac{\mu^c[s,u^{c\star}]}{\ones^\top \mu^c[s,\cdot]}F^{c,s}_{u^{c\star}}(\mu) \notag\\
		& \quad \quad \quad   - \sum_{v\in \UDc \setminus \{u^{c\star}\}}\frac{\mu^c[s,v]}{\ones^\top \mu^c[s,\cdot]}F^{c,s}_v(\mu)\notag\\
		& =  -(F^{c,s}_{u^{c\star}}(\mu) -F^{c,s}_u(\mu))\notag \\
		& \quad   +  \sum_{v\in \UDc \setminus \{u^{c\star}\}}\frac{\mu^c[s,v]}{\ones^\top \mu^c[s,\cdot]}(F^{c,s}_{u^{c\star}}(\mu) -F^{c,s}_v(\mu))\notag\\
		\notag\\
		& \leq - \min_{v\neq u^{c\star}}\{F^{c,s}_{u^{c\star}}(\mu) -F^{c,s}_v(\mu)\} \\
		& \quad + \frac{\alpha}{\ones^\top \mu^c[s,\cdot]} \max_{v\in \UDc \setminus \{u^{c\star}\}}\{F^{c,s}_{u^{c\star}}(\mu) -F^{c,s}_v(\mu)\}.\notag
	\end{align}}
	Notice that, for sufficiently small $\alpha^\prime>0$, there exists $k^{c,s}_1>0$ such that $\min_{v\in \UDc \setminus \{u^{c\star}\}}\{F^{c,s}_{u^{c\star}}(\mu) -F^{c,s}_v(\mu)\} > k_1^{c,s}$ for all $s\in \Scal^c$ and all $\mu \in D_{\alpha^\prime}(\mu^\star)$. Similarly, there is $k^{c,s}_2>0$ such that $\max_{v\in \UDc \setminus \{u^{c\star}\}}\{F^{c,s}_{u^{c\star}}(\mu) -F^{c,s}_v(\mu)\} < k^{c,s}_2$ for all $s\in \Scal^c$ and all $\mu \in D_{\alpha^\prime}(\mu^\star)$ because it is the maximum of a continuous function in a compact set. It follows from \eqref{eq:aux_avg_payoff} that $\hat{F}^{c,s}_u(\mu)< -k^{c,s}_1+\alpha k^{c,s}_2$ for all $u \in \UDc \setminus \{u^{c\star}\}$ and all $\mu \in D_{\alpha^\prime}(\mu^\star)$. 
	Therefore, choosing sufficiently small $\alpha_1^c$ such that $0< \alpha_1^c < \alpha^\prime$, it follows that $\hat{F}^{c,s}_u(\mu)< 0$ for all $s\in \Scal^c$, all $u \in \UDc \setminus \{u^{c\star}\}$ and all $\mu \in D_{\alpha_1^c}(\mu^\star)$. Since the excess payoff protocol is separable, it is also sign preserving  (see \cite[Exercise~5.5.6]{Sandholm2010}), i.e., $\sign(\tau^{c,s}_u(\hat{F})) = \sign(\max(0,\hat{F}^{c,s}_u(\mu)))$, therefore $\tau^{c,s}_u(\hat{F}^{c,s}(\mu))= 0$ for all $s\in \Scal^c$, all $u \in \UDc \setminus \{u^{c\star}\}$,  and all $\mu \in D_{\alpha_1^c}(\mu^\star)$. As a result, $f_{s,u^{c\star}}^{c,r}(\mu) = \tau^{c,s}_{u^{c\star}}(\hat{F}^{c,s}(\mu))\sum_{u \in \UDc \setminus \{u^{c\star}\}} \mu^c[s,u] \geq 0$. Furthermore, $\hat{F}^{c,s}_{u^{c\star}}(\mu) \geq 0$ for all $s\in \Scal^c$ and all $\mu \in D_{\alpha_1^c}(\mu^\star)$ with equality if and only if $\sum_{s\in \Scal^c}\mu^c[s,u^{c\star}] = m^c$, therefore $\tau^{c,s}_u(\hat{F}^c(\mu)) \geq 0$ with equality if and only if $\sum_{s\in \Scal^c}\mu^c[s,u^{c\star}] = m^c$.  As a result, $f_{s,u^{c\star}}^{c,\mathrm{r}}(\mu) = \tau^{c,s}_{u^{c\star}}(\hat{F}^{c,s}(\mu))\sum_{u \in \UDc \setminus \{u^{c\star}\}} \mu^c[s,u] \geq 0$ is null if and only if  $\sum_{s\in \Scal^c}\mu^c[s,u^{c\star}] = m^c$, which establishes statement~(ii) for class $c$.
		
	Third, we consider any class $c\in [C]$ that uses a pairwise comparison revision protocol. Since $\mu\in D_{\alpha}(\mu^\star)$ is a bounded neighborhood of $\mu^\star$ for any $\alpha>0$ and  $F_{u^{c\star}}^{c,s}(\mu^\star) > F_{v}^{c,s}(\mu^\star)$ for all $v\in \UDc \setminus \{u^{c\star}\}$ and all $s\in \Scal^c$, by continuity, for a sufficiently small $\alpha_1^c$ such that $\alpha_1^c > 0$, $\mu\in D_{\alpha_1^c}(\mu^\star) \implies \left(F_{u^{c\star}}^{c,s}(\mu) > F_{v}^{c,s}(\mu),  \forall v\in \UDc \setminus \{u^{c\star}\} \; \forall s\in \Scal^c \right)$. It follows that for all $\mu\in D_{\alpha_1^c}(\mu^\star)$ all $v\in \UDc \setminus \{u^{c\star}\}$  and all $s\in \Scal^c$ $\tau^c_{v,u^{c\star}}(F^{c,s}(\mu)) > 0$ and $\tau^c_{u^{c\star},v}(F^{c,s}(\mu)) = 0$. As a result, 
	\begin{equation}\label{eq:pairwise_cmp_signa_fr_strict}
		f_{s,u^\star}^{c,\mathrm{r}}(\mu) = \sum_{v\in \UDc \setminus \{u^{c\star}\}} \mu^c[s,v]\tau^c_{vu^{c\star}}(F^{c,s}(\mu)) \geq 0.
	\end{equation}
	Furthermore, \eqref{eq:pairwise_cmp_signa_fr_strict} holds with equality if and only if $\sum_{v\in \UDc \setminus \{u^{c\star}\}}\mu^c[s,v] = 0$. Finally, since $C$ is finite, one can choose $\alpha_1 := \min_{c\in[C]} \alpha^c_1$ to conclude the proof.
\end{proof}

\begin{proposition}\label{prop:dini_state_strict}
	Let $\{\mu(t)\}_{t\geq0}$ be a solution trajectory to \eqref{eq:ODE_mu_ev} according to an imitative, separable excess payoff, or pairwise comparison revision protocol. Then, $D^+||\mu^c[\cdot,u^{c\star}](t)-\mu^{c\star}[\cdot,u^{c\star}]||_1 \leq ||f^{c,\mathrm{r}}_{\cdot,u^{c\star}}(\mu(t))||_1\;$ for all $t\geq 0$ and all $c\in[C]$.
\end{proposition}
\begin{proof}
	From \eqref{eq:ODE_mu_ev} one can write
	\begin{equation}\label{eq:strict_linear_sys_with_rev_input}
		\dot{\mu}^c[\cdot,u^{c\star}](t) = Q \mu^c[\cdot,u^{c\star}](t) + f^{c,\mathrm{r}}_{\cdot,u^{c\star}}(\mu(t)),
	\end{equation}
	where $Q$ is the generator of the Markov process whose transition kernel is $\phi^{c,u^{c\star}} : \Scal^c \to \Pcal(\Scal^c)$ defined by $\phi^{c,u^{c\star}}(s|s^\prime) =  \sum_{a^\prime \in \Acal^c}\phi^c(s|s^\prime,a^\prime)u^{c\star}(a^\prime|s^\prime)$. Analyzing \eqref{eq:strict_linear_sys_with_rev_input} as a linear system with an exogenous input $f^{c,\mathrm{r}}_{\cdot, u^{c\star}}(\mu(t))$, one may conclude that it is weakly infinitesimally contracting~\cite[Definition~4.2]{Bullo2024}. Therefore, comparing the solution trajectory $\{\mu(t)\}_{t\geq0}$ with the degenerate solution trajectory that remains at $\mu^\star$ and using \cite[Theorem~3.16]{Bullo2024}\footnote{Although \cite[Theorem~3.16]{Bullo2024} is stated for strongly contracting systems the proof can be reused with minimal changes to establish an analogous result for weakly contracting systems.} one obtains $D^+||\mu^c[\cdot,u^{c\star}](t)-\mu^{c\star}[\cdot,u^{c\star}]||_1 \leq ||f^{c,\mathrm{r}}_{\cdot,u^{c\star}}(\mu(t))||_1$ for all $t\geq 0$.
\end{proof}

Now, let $\{\mu(t)\}_{t\geq0}$ be a solution trajectory to \eqref{eq:ODE_mu_ev}. The candidate Lyapunov function is 
\begin{equation}\label{eq:lyapunov_strict}
	\begin{split}
		V(\mu(t)) &= \sum_{c\in[C]} ||\mu^c[\cdot,u^{c\star}](t)-\mu^{c\star}[\cdot,u^{c\star}]||_1 \\
		&+ K\!\! \sum_{c\in [C]} \sum_{v\in \UDc \setminus \{u^{c\star}\}}\!\!\!\!\!||\mu^c[\cdot,v](t) - \mu^{c\star}[\cdot,v]||_1,
	\end{split}
\end{equation}
	where $K>0$ is a constant to be chosen.
	By the subadditivity of the $\limsup$ operator, it follows that $D^+V(\mu(t)) \leq \sum_{c\in [C]}D^+ ||\mu^c[\cdot,u^{c\star}](t)-\mu^{c\star}[\cdot,u^{c\star}]||_1 + K \sum_{c\in [C]} D^+\sum_{v\in \UDc \setminus \{u^{c\star}\}}||\mu^c[\cdot,v](t)||_1$. Since the second term in \eqref{eq:lyapunov_strict} is continuous and differentiable, it follows from statement~(i) of Lemma~\ref{lemma:prop_Dini} that its upper right Dini derivative is the usual derivative, so 
	\begin{equation*}
	\begin{split}
			&D^+ \!\!\!\!\!\!\sum_{v\in \UDc \setminus \{u^{c\star}\}}||\mu^c[\cdot,v](t)- \mu^{c\star}[\cdot,v]||_1 \\
			= &  \sum_{s\in \Scal^c} \sum_{v\in \UDc \setminus \{u^{c\star}\}} \dot{\mu}^c[s,v](t) \\
			= & \sum_{s\in \Scal^c} \sum_{v\in \UDc \setminus \{u^{c\star}\}} f^{c,\mathrm{r}}_{s,v}(\mu(t)) =  -\sum_{s\in \Scal^c}   f^{c,\mathrm{r}}_{s,u^{c\star}}(\mu(t)),
	\end{split}
\end{equation*}
where the last two equalities are due to $\sum_{s\in\Scal^c}f^{c,\mathrm{d}}_{s,v}(\mu(t)) = 0$ for all $v\in \UDc$ and $\sum_{u\in \UDc}f^{c,\mathrm{r}}_{s,u}(\mu(t)) = 0$. As a result, from statement~(i) of Proposition~\ref{prop:strict_propoerties_rev} and Proposition~\ref{prop:dini_state_strict}, it follows that if $\mu(t) \in D_{\alpha_1}(\mu^\star)$ and choosing $K>1$
\begin{equation*}
	\begin{split}
				\begin{split}
				\!\!D^+V(\mu(t)) &\leq  \!\!\!\sum_{c\in [C]}\!||f^{c,\mathrm{r}}_{\cdot,u^{c\star}}(\mu(t))||_1 \! -K\!\!\!\sum_{c\in[C]}\sum_{s\in \Scal^c} f^{c,\mathrm{r}}_{s,u^{c\star}}(\mu(t))\! \!\\
				& \leq  \!\!\!\sum_{c\in [C]}\! \sum_{s\in \Scal^c} \!\! f^{c,\mathrm{r}}_{s,u^{c\star}}(\mu(t)) \! -\!K\!\!\!\sum_{c\in[C]}\sum_{s\in \Scal^c}f^{c,\mathrm{r}}_{s,u^{c\star}}(\mu(t))\! \!\\
				&\leq   -(K-1) \sum_{c\in[C]}\sum_{s\in \Scal^c} f^{c,\mathrm{r}}_{s,u^{c\star}}(\mu(t)) \leq 0.
			\end{split}
	\end{split}
\end{equation*}
Notice that the set $\Omega := \left\{\mu \in X : V(\mu) \leq \alpha_2 \right\}$ is compact for any $\alpha_2>0$ and, for sufficiently small $\alpha_2>0$,  $\Omega \subset D_{\alpha_2}(\mu^\star)$. As a result, $D^+V(\mu(t)) \leq 0$ if $\mu(t) \in \Omega$ and, by statement~(ii) of Lemma~\ref{lemma:prop_Dini}, $\Omega$ is a positive invariant set. By a simple generalization of LaSalle's Invariance Principle \cite[Theorem~3.16]{Bullo2024} every trajectory approaches the largest invariant set in $E = \{\mu \in \Omega : D^+V(\mu) = 0\}$. For details on the generalization, one can refer to the discussion on \cite[Chap.~A.7]{Bullo2024} or check that the proof of LaSalle's Invariance Principle in \cite[Theorem~4.4]{Khalil2002} holds in this setting as well. From statements (i) and (ii) in Proposition~\ref{prop:strict_propoerties_rev}, we have that
\begin{equation*}
	\begin{split}
			& D^+V(\mu) = 0\\
			\implies & -(K-1) \sum_{c\in[C]}\sum_{s\in \Scal^c} f^{c,\mathrm{r}}_{s,u^{c\star}}(\mu(t)) = 0 \\
			\implies &  f^{c,\mathrm{r}}_{s,u^{c\star}}(\mu) = 0 \; \forall c\in [C]\;\forall s \in \Scal^c  \\
			\implies & \sum_{c\in [C]}\sum_{s\in \Scal^c}\sum_{v\in \UDc \setminus \{u^{c\star}\}}\mu^c[s,v] = 0.
	\end{split}
\end{equation*}
Therefore, by Assumption~\ref{ass:noise}, the largest invariant set contained in $E$ is  $\{\mu^\star\}$, thus, $\mu^\star$ is locally asymptotically stable, which concludes the proof.